%% file: Arxiv.tex
\documentclass[twocolumn]{cls/autartPreprint}    % Enable this line and disable the 
\usepackage{graphicx}          % Include this line if your 
\usepackage{amsmath}
\usepackage[dvipsnames]{xcolor}
\usepackage{amssymb}
\usepackage{pifont}% http://ctan.org/pkg/pifont

\usepackage{./cls/HT_command}
 
\def \pics {./pics/pdf/} 
\usepackage{algorithm}

\usepackage{tikz}
\usepackage{amsmath}
\usepackage{tikz}
\usepackage{mathdots}
\usepackage{xcolor}
\usepackage{soul}
\usepackage{cancel}
\usepackage{color}
\usepackage{siunitx}
\usepackage{array}
\usepackage{multirow}
\usepackage{amssymb}
\usepackage{tabularx}
\usepackage{pgfplots}
\usepgfplotslibrary{groupplots}
\usetikzlibrary{plotmarks}
\pgfplotsset{compat=1.18} 
\usepgfplotslibrary{patchplots}
\newtheorem{assumption}{Assumption}

\usepackage{mathtools} %for mathclap

\definecolor{darkgreen}{HTML}{1e8f37}
\begin{document}
	
	\begin{frontmatter}
		
		\title{Reducing real-time complexity via sub-control Lyapunov functions: from theory 
			to experiments
			\thanksref{footnoteinfo}} % Title, preferably not more 
		
		\thanks[footnoteinfo]{Corresponding author H.~T.~Do.}
		
		\author[lcis]{Huu-Thinh Do}\ead{huu-thinh.do@lcis.grenoble-inp.fr},
		\author[MeI-UdU]{Franco Blanchini}\ead{blanchini@uniud.it}
		,
		\author[MeI-UdU]{Stefano Miani}\ead{miani.stefano@uniud.it}
		,
		\author[lcis]{Ionela Prodan}\ead{ionela.prodan@lcis.grenoble-inp.fr},    
		
		% e-mail address 
		% \author[Baiae]{Publius Maro Vergilius}\ead{vergilius@culture.ir}  % (ead) as shown
		
		\address[lcis]{Univ. Grenoble Alpes, Grenoble INP$^\dagger$, LCIS, F-26000, Valence, France.
			\\$^\dagger$ Institute of Engineering and Management University Grenoble Alpes.}  % Please supply                                              

		\address[MeI-UdU]{Dipartimento di Matematica e Informatica,
			Universit\`a di Udine, 33100 Udine, Italy.}

		% \address[Rome]{Senate House, Rome}             % full addresses
		% \address[Baiae]{The White House, Baiae}        % here.

		                             % keyword list or with the 
		% help of the Automatica 
		% keyword wizard

		\begin{abstract}                          % Abstract of not more than 200 words.
			The techniques to design control Lyapunov functions (CLF), along with a proper stabilizing feedback, possibly in the presence of constraints, often provide control laws that are too complex for proper implementation online, especially when an optimization problem is involved. In this work, we show how to acquire an alternative, computationally attractive feedback. Given a nominal CLF and a nominal state feedback, we say that a different positive definite function is a Sub-control Lyapunov function (SCLF) if its Lyapunov derivative is negative-definite and bounded above by the Lyapunov derivative of the nominal function with the nominal control. It turns out that if we consider a family of basis functions, then a SCLF can be computed by linear programming, with an infinite number of constraints. The idea is that although the offline computational burden to achieve the new controller and solve the linear program is considerable, the online computational burden is drastically reduced. Comprehensive simulations and experiments on drone control are conducted to demonstrate the effectiveness of the study.
			% The effectiveness of the framework will be examined via both simulations and experiments on drone control. 
		\end{abstract}
	\end{frontmatter}
	
	\setlength\abovedisplayskip{3.2pt}
	\setlength{\belowdisplayskip}{3.2pt}

	\section{Introduction}
	The availability of powerful software packages for optimization \cite{andersson2019casadi,stripinis2022directgo,cvx}, allows designing controllers with high performance. The clearest example is the ever-growing popularity of Model Predictive Control (MPC), where the stabilizing feedback policy is computed implicitly via an optimization problem, handling directly operating constraints and balancing between performance and actuator exploitation \cite{blanchini2003suboptimal,chen1998quasi}. Additional features, such as safety or robustness, can also be added by properly modifying the cost and constraints of the program \cite{wu2019control,zeng2021safety}.
	However, these sophisticated controllers often necessitate a large amount of processing power 
	real-time, limiting their application to systems with ample computational capacity or slow dynamics.
	
	% but they are frequently prohibitively expensive to implement, limiting their use to systems with either slow dynamics or ample computational capability.

	% \begin{figure}[htbp]
		%     \centering
		%     \resizebox{0.247\textwidth}{!}{\input{pics/tikz/overallScheme}}
		%     \caption{Control offline synthesis and online deployment with sub-control Lyapunov functions.}
		%     \label{fig:overallScheme}
		% \end{figure}
	
	To face this problem, in \cite{bemporad2002explicit},
	an explicit expression of the optimal control is adopted as a look-up table,
	to be used online. However, its implementation becomes increasingly intractable with respect to the system's dimensions and the length of the prediction horizon \cite{kvasnica2013complexity}. An alternative direction to attain online operability for these policies is to learn the implicit control law with machine learning using neural networks as a function approximator, such as in \cite{chen2018approximating,drgovna2022differentiable,mukherjee2022neural} or \cite{karg2020efficient}, to name a few. This synthesis can be viewed as a supervised learning problem where the feedback variables are the features labeled with the input decisions made by such an ``expert" control. Although the strategy significantly reduces the computational footprint, there is still room for improvement in terms of fitting performance, theoretical guarantee, and training data characterization, especially when large-scale systems come into play \cite{chen2022large,drgovna2018approximate}.
	
	From a control theoretic viewpoint, stability can be ensured if there exists a control Lyapunov function (CLF) for the closed-loop dynamics. If successfully synthesized, the function provides a stability certificate, which can be exploited to derive efficient control laws, for instance
	adopting Sontag's formula \cite{sontag1989universal,yamashita2022asymptotic} or a min-norm control \cite{freeman1996inverse}. 
	
	A considerable effort has been devoted to synthesizing CLF. For instance, generally without any pre-constructed controller, piecewise continuous CLF can be numerically generated via solving a semi-definite program (SDP) \cite{lavaei2023systematic} or a linear program (LP) \cite{steentjes2020construction,julian1999parametrization,fiacchini2015computation,RavSan19,Lazar09}. 
	%Although offering appealing and complete analytical guarantees, the approaches are hardly practical in real time, even with modern hardware, due to the so-called \textit{point location problem} induced by the state-space triangulation. 
	For a complete survey on numerical methods for CLF synthesis, we refer to 
	\cite{giesl2015review}. 
	
	In this work, we assume that a CLF, $\Psi(x)$,
	and a feedback control 
	$\omega(x)$, are available
	(possibly from optimal design) but
	that these functions are unsuitable
	for real-time implementation.
	The idea we pursue is that of replacing $\Psi(x)$ and $\omega(x)$
	by a simpler $V(x)$,  we call Sub-control Lyapunov function, along with a feedback $\bar u(x)$
	amenable for real-time
	implementation.

	Our solution is based on the parameterization of the CLF with a set of basis functions (see \cite{giesl2014revised,RavSan19,do2023lp} and others). The non-positivity constraints of the function's time derivative become linear with respect to the variable parameters. Since these conditions are imposed continuously over the state space, we need to cope with infinitely many constraints
	\cite{johansen2000computation}. 
	This approach was validated 
	in our previous work \cite{do2023lp}, although the optimization problem was relaxed by randomly sampling the state space.
	
	% The goal of this paper is that of proposing
	% a theoretical framework. Specifically, we:
	% \begin{itemize}
		%     \item formalize the concept of Sub-Control Lyapunov Functions (SCLF), which retains the
		%     optimal controller 
		%     as possible feedback,
		%     but it also admits
		%     alternative low-cost feedback laws amenable
		%     for implementation;
		%     \item propose a comprehensive computational procedure to generate an SCLF based on the cutting-plane technique for semi-infinite linear programming;
		%     \item validate the proposed framework with illustrative examples and experimental tests.
		% \end{itemize}
	
	In this context of CLF computation, our goal is to provide a theoretical framework taking into account the performance of the deduced control. Specifically, we:
	\begin{itemize}
		\item formalize the concept of Sub-Control Lyapunov Functions (SCLF), 
		which retains the
		optimal controller 
		as possible feedback,
		but it also admits
		alternative low-cost feedback laws amenable
		for implementation;
		\item introduce a performance-based formulation to generate such an SCLF, describing a Semi-Infinite Linear Program (SILP) and propose an algorithm to solve the problem based on cutting-plane method;
		% \item propose a computational procedure to generate an SCLF based on the cutting-plane technique to solve the problem;
		\item validate the proposed framework with illustrative examples and experimental tests.
	\end{itemize}
	
	% \cite{blanchini2003suboptimal,do2023lp,isidori1989nonlinear,ross2006issues,nguyen2022improved}
	% ILP \cite{hettich1993semi,polak2012optimization,reemtsen2013semi}
	% \cite{mariano2020asymmetric}
	% \cite{hu2006stability,milani2002piecewise,lu2011synthesis}
	
	% CLF related \cite{girard2019lyapunov,gong2022constructing,fitzsimmons2024neural,masti2023counter,ballaben2024lyapunov}
	
	% \cite{lekic2017controlling}

	% \cite{ahmed2020automated}

	% In Section \ref{sec:sCLF_def} we define a SCLF, proving its basic properties, and we show how to derive low-cost stabilizing controllers. We show how to achieve local optimality for linearizable systems, and we evaluate the closed-loop performance loss. 
	% In Section  \ref{sec:computation_sCLF} 
	% we describe a computational approach for searching a SCLF via Semi-Infinite Linear Program (SILP). Section \ref{sec:numericalExp} shows that the proposed framework can be applied for nonlinear systems of non-trivial size with simulation studies. The proposed method is experimentally validated via a quadcopter position control application. Section \ref{sec:Conclusion} concludes and discusses future work.

	In Section  \ref{sec:computation_sCLF} 
	we describe a computational approach to search for an SCLF via an SILP. Section \ref{sec:numericalExp} shows that the framework can be applied for nonlinear systems of non-trivial size with simulation studies. The method is also experimentally validated via a quadcopter control application. Section \ref{sec:Conclusion} concludes and discusses future work.

	\textit{Notation:} For $x,y\in\R^n$, $x_i$ is its $i$th entry,  $|x|\leq|y|$ implies $|x_i|\leq|y_i|\ \forall i$ and $\|x\|$ denotes the Euclidean norm.  With $x^*\in\R^n$, $\epsilon>0$, $\mathcal{B}_\epsilon(x^*)\triangleq\{x\in\R^n:\|x-x^*\|\leq \epsilon\}$. The function $\underset{|u|\leq u_{max}}{\text{sat}(w)} $ returns the standard vector saturation with its $i$th component computed as $\max(\min(u_{i,max},w_i),-u_{i,max})$.  For an integer $\ell$, $\mathcal{N}(\ell,a,b)$ denotes the set of $\ell$ real numbers evenly sampled in the interval $[a,b]$. $\mathrm{co}(\cdot)$ is the convex hull operator. diag$(\cdot)$ returns a block matrix with its entries diagonally arranged. All-ones vector is denoted as $\mathbf{1}$.
	
	\section{Sub-Control Lyapunov Functions}
	\label{sec:sCLF_def}
	\subsection{Definition and basic properties}
	Consider the system
	\begin{equation} \label{sys}
		\dot x = A(x) + B(x) u,\,\, u\in\mathcal{U},
	\end{equation}
	with $x\in\R^n,u\in\R^m$ denoting the state and input vector, respectively. $\mathcal U$ is the input constraint set.
	We consider the  control problem of stabilizing the system at the equilibrium point $x=0,u=0$.
	
	We call a control Lyapunov function (CLF), associated with the control $\omega(x):\R^n\rightarrow \mathcal{U}$,
	a locally Lipschitz positive definite function
	$\Psi(x): \R^n\rightarrow \R$  with the property:
	\begin{equation} \label{ineq}
		D^+\Psi(x,\omega(x))+ g(x,\omega(x)) = 0
	\end{equation}
	where $g(x,u)$ is locally Lipschitz and positive for non-zero $(x,u)$, and where we have denoted by 
	\begin{equation}
		\begin{aligned}
			&D^+\Psi(x,\omega(x)) \\ &=\limsup_{h\rightarrow 0^+}\frac{\Psi(x+h(A(x)+B(x)\omega(x)))-\Psi(x)}{h},
		\end{aligned}
		\label{limsup}
	\end{equation}
	the generalized directional derivative. In the smooth case of $\Psi(x)$, one clearly has:
	\begin{equation}
		\begin{aligned}
			D^+\Psi(x,\omega(x)) & = \nabla \Psi(x)\left (A(x)+B(x)\omega(x)  \right ).
		\end{aligned}
	\end{equation}
	We consider the generalized derivative, since $\Psi(x)$ could be non-differentiable.
	
	Note that function $g(x,u)$ having the property \eqref{ineq} can be  associated with the performance index:
	\begin{equation} \label{cost}
		J(x(0),u(\cdot)) = \int_0^\infty g(x(t),u(t))\dd t
	\end{equation}
	while $\Psi(x)$ is  the cost--to--go  function, solution of the Hamilton-Jacobi-Bellman equation:
	\begin{equation} \label{HJB}
		\begin{aligned}
			&\min_{u \in \mathcal{U}}\left [ D^+\Psi(x,u) + g(x,u) \right ] \\&= D^+\Psi(x,\omega(x)) + g(x,\omega(x))=0
		\end{aligned}
	\end{equation}
	where we have denoted by $\omega(x)$ the minimizer control.
	
	Finding $\Psi(x)$ and a an optimal control law   $\omega(x)$, often requires brute force 
	computation via numerical approaches, moreover they are often unsuitable for online implementation.
	To arrive at an approximation scheme, we introduce the following definition.
	\begin{definition}
		\label{def:sCLF_def}
		Given a polynomially bounded nominal control Lyapunov function (CLF) $\Psi(x)$ equipped with the control
		$\omega(x)$ we say that the polynomially bounded positive definite function $V(x)$ is a global sub-control Lyapunov function (SCLF) of $\Psi(x)$, if
		\be
		D^+V(x,\omega(x)) \leq D^+\Psi(x,\omega(x)) = -  g(x,\omega(x)),~\forall x.
		\label{eq:sCLF_def}
		\ee
		It is a local SCLF if the inequality holds for $V(x) \leq \kappa$, for some $\kappa >0$.
	\end{definition}
	
	There is an important relation between $V(x)$ and $\Psi(x)$.
	\begin{prop} \label{VgeqPsi}
		If $~V(x)$ is a SCLF of $~\Psi(x)$, then $V(x) \geq \Psi(x)$.
	\end{prop}
	\begin{pf}
		Let $x_0$ be an initial condition and consider the converging trajectory $\chi(t)$,
		such that $\chi(0)=x_0$,
		associated with the control $\omega(x)$.
		By integrating, we get
		$$
		\begin{cases}
			V(x_0) - V(\chi(\infty)) = - \int_0^\infty~~ D^+V(\chi(t),\omega(\chi(t))) \dd t &\\
			\Psi(x_0) - \Psi(\chi(\infty)) = - \int_0^\infty~~ D^+\Psi(\chi(t),\omega(\chi(t))) \dd t .&
		\end{cases}
		$$
		As the control is stabilizing, $\chi(\infty)=0$. Hence, $V(x_0) \geq \Psi(x_0)$.
		Since $x_0$ is arbitrary, the claim follows.\QEDA
	\end{pf}
	
	In the sequel, for practical implementation, we consider smooth candidate SCLF. 
	%We have the property that the set of all smooth SCLF, given $\Psi(x)$ and $\omega(x)$ is convex and this will enable us to use linear programming as a synthesis machinery.

	% \begin{rem} There is a hidden technical issue. The regular derivative has the additive property
		% $d/dt (f(t)+g(t)= d/dt f(t)+d/dtg(t)$. This is not true for the $\limsup$ derivative.
		% \textcolor{red}{I do not understand this remark.}
		% \end{rem}
	
	\subsection{Suboptimal control laws from SCLFs}
	Now we show how
	low-cost controllers can
	be associated with a smooth SCLF (with a performance loss).
	As can be seen from \eqref{ineq}, by integrating with $x(0)=x_0$, one can write the optimal performance as:
	\begin{equation}
		J(x_0,\omega(x))=- \int_0^\infty D^+\Psi(x,\omega(x)) \dd t= \Psi(x_0),
		\label{bound}
	\end{equation}
	If the control $u=\omega(x)$ is  unsuitable
	for real time implementation, we could replace
	$\Psi(x)$ by a smooth SCLF $V(x)$ defined  as in \eqref{eq:sCLF_def}, and adopt the following control:
	\begin{equation} \label{control}
		\bar u(x) = \arg\min_{u \in \mathcal{U}} \nabla V(x)  [A(x) + B(x) u  ] + g(x,u).
	\end{equation}
	It is fundamental to note that problem \eqref{control} is feasible because
	$u=\omega(x)$, the optimal, is a feasible solution.
	The next theorem explains how
	this control performs.
	\begin{thm} Assume that the nominal $\Psi(x)$ satisfies \eqref{ineq}. Consider a SCLF $V(x)$.
		Then control $\bar u(x)$ as in \eqref{control} is stabilizing. Moreover, given the cost function $g(x,u)$ 
		and the initial condition $x(0)=x_0$, the performance is upper bounded by $V(x_0)$.
		\label{thm:guaran}
	\end{thm}
	
	\begin{pf}
		We have that
		\begin{equation}
			\begin{aligned}
				& \dot V(x,\bar u(x)) +
				g(x,\bar u(x))
				= \nabla V(x) [A(x) + B(x) \bar u(x)] \\ &+ g(x,\bar u(x)) 
				\leq \underbrace{\nabla V(x) [A(x) + B(x) \omega(x)]}_{D^+V(x,\omega(x))}
				+ g(x,\omega(x)) \\
				&\leq D^+ \Psi(x,\omega(x)) + g(x,\omega(x)) = 0. \end{aligned}
			\nonumber
		\end{equation}
		The first inequality is due to the fact that $\bar u(x)$ is the minimizer of the optimization problem \eqref{control}, the second equality follows from the definition
		of a SCLF
		\eqref{eq:sCLF_def}. Adopting the control $\bar u(x)$, the time derivative of $V(x)$ is
		\begin{equation}
			\dot V (x,\bar u(x)) =\nabla V(x) [A(x) + B(x) \bar u(x)] 
			\leq  -g(x,\bar u(x)),
			\label{eq:dev_V_minus_g}
		\end{equation}
		which implies asymptotic stability because $g(x,u)$ is positive definite, hence $V(\infty)=\underset{t\rightarrow \infty}{\lim}V(x(t))=0$. Then, integrating \eqref{eq:dev_V_minus_g} leads to:
		\begin{equation}
			\int_0^\infty~~g(x,\bar u(x)) \dd t \leq V(x_0) - V(\infty) = V(x_0) .
			\label{eq:upper_cost}
		\end{equation}
		\QEDA
		
	\end{pf}
	
	The computational complexity of the  optimization problem \eqref{control} (to be solved in real-time), 
	is very low and we can even provide  a simple explicit solution in the case of a quadratic cost $g(x,u)$ and a box-type set $\mathcal{U}$ in \eqref{sys}.
	
	\begin{prop}
		\label{prop:sat_control_R}
		
		Consider a quadratic cost:
		\begin{equation}
			g(x,u) = x^\top Q x + u^\top R u,
			\label{eq:quad_cost}
		\end{equation}
		with a positive definite diagonal weighting matrix $R=\text{diag}(r_{1},r_{2},...,r_{m})$ and a constraint set:
		\be
		\mathcal{U} = \left\{u  \in\R^m:|u| \leq u_{max}\right\}.
		\label{eq:box_U}
		\ee 
		Then, the minimizer of \eqref{control} can be explicitly computed as:
		\be 
		\label{eq:closedForm_control}
		\bar u(x) =\underset{|u| \leq u_{max}}{\text{sat}}\left( -\textstyle{\frac{1}{2}} R^{-1} B(x)^\top \nabla V(x)^\top
		\right)
		\ee 
	\end{prop}

	\begin{pf}
		Let $z=(\textstyle{\frac{1}{2}} \nabla V(x) B(x))^\top$ and $y=R^{-1}z$, then the problem \eqref{control} can be rewritten as:
		\begin{equation}
			\begin{aligned}
				& \bar u(x) = \arg\min_{u\in\mathcal U}~ 2z^\top u+ u^\top R u, \\
				& = \arg\min_{u\in\mathcal U}~ (u+y)^\top R (u+y) - y^\top R y.
			\end{aligned}
			\label{eq:sat_proof1}
		\end{equation}
		With the set $\mathcal{U}$ in \eqref{eq:box_U}, \eqref{eq:sat_proof1} is decoupled
		\begin{equation}
			\bar u(x) = \underset{|u| \leq u_{max}}{\arg\min}~ \sum_{i=1}^m r_i(u_i+y_i)^2.
			\label{eq:sat_proof2}
		\end{equation}
		For each component $\bar u_i(x)$ of $\bar u(x)$, the minimizer of \eqref{eq:sat_proof2} is:
		$  
		\bar u_i(x) = \underset{|u_i| \leq u_{i,max}}{\text{sat}}(-y_i),
		\label{eq:sat_proof3}
		$
		or equivalently, \eqref{eq:closedForm_control}.
		%  \begin{equation}
			% \bar u(x)= \underset{|u| \leq u_{max}}{\text{sat}}(-\textstyle{\frac{1}{2}}R^{-1}B(x)^\top \nabla V(x)^\top)\end{equation}
		\QEDA
	\end{pf}
	For non-diagonal positive definite $R$, Proposition \ref{prop:sat_control_R} does not hold.
	Still, one obtains a solution for the optimization problem as follows.
	\begin{prop}
		Let $u \in \mathcal{U}$ a polytope. Then problem (9)
		with cost (12), reduces to a minimum distance problem.
	\end{prop}
	\begin{pf}
		Denoting by $z = \frac{1}{2} \left(\nabla V (x)B(x) \right )^\top $, $\bar u(x)$ is the
		minimizer of
		$$\bar u(x) = \mbox{arg}\min_{u\in \mathcal{U}} 2z^\top u + u^\top Ru$$
		
		Let $v = R^{\frac{1}{2}} u$ with $v \in \mathcal{V} \triangleq R^{\frac{1}{2}} \mathcal{U}$, the problem is
		equivalent, after completing the squares, to:
		
		$$\begin{array}{l} \underset{v \in \mathcal{V}}{\min}\left\{
			-z^\top R^{-1} z + z^\top R^{-1} z + 2z^\top R^{-\frac{1}{2}} v + v^\top v \right \}\\
			=\underset{v \in \mathcal{V}}{\min}\left\{ -z^\top R^{-1} z+\left( R^{-\frac{1}{2}}z+v\right )^\top \left( R^{-\frac{1}{2}}z+v)^\top  \right ) \right \} \\
			=\min_{v \in \mathcal{V}}\left\{ -z^\top R^{-1} z+\| R^{-\frac{1}{2}}z+v \|^2\right \}.
		\end{array}
		$$
		So $v = R^{\frac{1}{2}} u$ is the point in $\mathcal{V}$ closest to $-R^{-\frac{1}{2}} z$.
		\QEDA
	\end{pf}
	
	\begin{rem}
		Solving the minimum distance problem 
		is straightforward. Moreover,
		if $z(x)=\frac{1}{2}\left(\nabla V (x)B(x)\right)^\top$ is continuous, then the minimizer function $\bar u(x)$ is continuous as well
		being the so called minimal selection
		\cite{AubCel84}.
	\end{rem}

	An essential property of a SCLF is that it admits, among other controllers the 
	true optimal $\omega(x)$ beside the simpler
	$\bar u(x)$ discussed above. We expect that
	$\bar u(x)$ is close to $\omega(x)$. We argue that we can achieve optimality
	close to the origin.
	Indeed, assume a cost of the form \eqref{eq:quad_cost} and the set of all control actions that ensure \eqref{eq:dev_V_minus_g}, or:
	\begin{equation}
		\label{ineqQR}
		\nabla V(x) [A(x) + B(x)  u]\leq  -x^\top Q x - u^\top R u, u \in \mathcal{U}    .
	\end{equation}
	Assume that, by means of linearization, we
	have determined an optimal desired
	gain $u=Kx$, the unconstrained local optimum, so that $\omega(x)\approx Kx$ close to $0$.
	
	Parameterize the control as
	$u = Kx + v \in \mathcal{U}$.
	Complete the square and rewrite
	\eqref{ineqQR}  as
	
	\begin{align}
		% \label{allcontrol1}
		&Kx+v \in \mathcal{U} :\left \|R^{\frac{1}{2}}\left[v+Kx+\frac{1}{2}R^{-1}B(x)^\top \nabla V(x) ^\top \right]\right \|^2
		\nonumber
		\\
		&\leq  -x^\top Q x  -\nabla V(x) A(x) 
		+\left \|\frac{1}{2}R^{-\frac{1}{2}}B(x)^\top \nabla V(x)^\top \right \|^2.
		\label{ineqquad}
	\end{align}
	By construction, \eqref{ineqQR} and \eqref{ineqquad}
	are always feasible. Hence, we can compute 
	the control as
	\begin{equation}
		\min \|v\| ~\text{s.t.}~~~Kx + v \in \mathcal{U}, \text{ and }
		\eqref{ineqquad}
	\end{equation}
	which is, again, a straightforward
	minimum norm problem with quadratic and 
	linear constraints. Since we assume that $\omega(x) \approx Kx$,
	and that $\omega(x)$ is a feasible control,
	in a neighborhood of $0$ we have the minimizer $v=0$ so that
	this new control will be virtually identical to $\omega(x)$.

	\subsection{Performance loss evaluation}
	
	Replacing $\Psi(x)$ by a SCLF 
	$V(x) \geq \Psi(x)$ (by Proposition \ref{VgeqPsi}) entails a performance loss which
	can be quantified by the index:
	\be 
	\mbox{loss}= \frac{V(x_0)}{\Psi(x_0)} \geq 1 .
	\label{eq:performanceLoss}
	\ee 
	Computing this index requires determining
	$\Psi(x)$ and $\omega(x)$, to impose the condition $D^+V(x,\omega(x)) \leq D^+\Psi(x,\omega(x)) =-g(x,\omega(x))$ to find $V(x)$ as we will see later.
	In practice, computing $\Psi(x)$, $D^+\Psi(x,\omega(x))$ explicitly and solving the $HJB$ optimality equation are affected by  numerical and even
	non--differentiability
	issues \cite{BarCapDol97}. Instead of solving the HJB equations, we can evaluate $\Psi(x)$
	and $\omega(x)$ in \eqref{ineq} implicitly, by
	considering the problem:
	\begin{subequations}
		\label{opti_infty}
		\begin{align}
			\Psi(x_0) &= \Psi_\infty(x_0) =&  \inf_{u(\cdot)\in \mathcal{U}}\int_{0}^{\infty}g(x(\tau),u(\tau))\mathtt{d}\tau
			\label{eq:cost_inf}
			\\
			& & 
			\dot x = A(x)+B(x)u , \quad  \label{eq:dyna_qih_inf}
			\\
			&&  x(0) = x_0.\quad\quad\quad\quad\;\;\,
			\label{x0_inf}
		\end{align}
	\end{subequations}
	% \begin{eqnarray}
		% \Psi(x_0) = \Psi_\infty(x_0) &=&  \inf_{u(\cdot)\in \mathcal{U}}\int_{0}^{\infty}g(x(\tau),u(\tau))\mathtt{d}\tau
		% \label{eq:cost_inf}
		% \\
		% & & \quad \quad \quad 
		% \dot x = A(x)+B(x)u , \label{eq:dyna_qih_inf}
		% \\
		% && \quad\quad\quad x(0) = x_0.
		% \label{x0_inf}
		% \end{eqnarray}
	
	The control $\omega(x_0)$ is the initial control value of the open-loop  problem
	\eqref{eq:cost_inf}--\eqref{x0_inf}.
	Solving the
	infinite horizon problem \eqref{eq:cost_inf}--\eqref{x0_inf} exactly
	may be difficult and, in practice, one has to approximate
	it by a finite (large) horizon 
	problem \cite{Ala06}:
	
	\begin{subequations}
		\label{eq:qihMPC}
		\begin{align}
			& \Psi_T(x_0) = \min_{u(\cdot)}{\int_{0}^{T}}g(x(\tau),u(\tau))\mathtt{d}\tau + \Phi(x(T)), \\
			& \quad \quad \quad \dot x = A(x)+B(x)u, \label{eq:dyna_qih}
			\\
			& \quad\quad\quad 
			u(\tau) \in \mathcal{U}, \tau \in [0,T], \\
			& \quad\quad\quad x(0) = x_0, \\
			& \quad\quad\quad x(T) \in \mathcal{P},
		\end{align}
	\end{subequations}
	where $T$ is the prediction horizon. $\mathcal{P}$ and $\Phi(\cdot)$ denote the controlled invariant terminal region and terminal cost function, respectively.  
	
	We now borrow ideas from Chen and Allgower \cite{chen1998quasi},
	to evaluate the infinite-time
	performance. Let us consider the following assumption.
	\begin{assumption}
		\label{inequality}
		The terminal cost  $\Phi(x)$ is smooth, positive definite and satisfies the HJB-like inequality
		\begin{align}
			% \label{nablaPhi}
			& \min_{u \in \mathcal{U}} \{\dot \Phi(x,u) +g(x,u)\} \nonumber \\
			&=   \min_{u \in \mathcal{U}} \left \{ \nabla \Phi(x) [A(x) + B(x)u] + g(x,u)\right \} \leq 0
			\label{ineqPsi}
		\end{align} 
		in the terminal set $\mathcal{P} = \{x:\Phi(x) \leq \kappa$ \}.
	\end{assumption}
	\begin{rem}
		For locally linearizable systems with the quadratic cost \eqref{eq:quad_cost}, we can take $\Phi(x)=x^\top S x$. Then $\Phi(x)$
		is a solution to \eqref{ineqPsi} if
		$S$ satisfies the Riccati inequality
		$$\bar A^\top S + S\bar A-S\bar BR^{-1}\bar B^\top S + Q \leq 0$$
		where $\bar A, \bar B$ are from the Jacobian linearization of \eqref{sys} at $x=0,u=0$.
		The value $\kappa>0$ 
		should be small enough
		so that \eqref{ineqPsi}
		remains valid in $\mathcal{P} = \{x:x^\top S x \leq \kappa \}$ \cite{chen1998quasi}.
	\end{rem}
	Then we have the following result.
	\begin{thm}
		Under Assumption \ref{inequality}
		consider the {\em finite horizon} model predictive
		scheme \eqref{eq:qihMPC}. This problem is recursively feasible. Assume that  
		the optimal control $\omega(x)$ is continuous and denote by  $\Psi_T(x)$ 
		its value function. We have
		\begin{equation}
			D^+\Psi_T(x,\omega(x)) \leq -g(x,\omega(x)),
			\label{ineq1}
		\end{equation}
		hence $\Psi_T(x_0)$ is a bound for the {\em infinite time} performance with $x(0)=x_0$, i.e.,
		\begin{equation}
			\int_0^\infty ~g(x,\omega(x)) \dd t \leq \Psi_T(x_0).
			\label{gbounds}
		\end{equation}
	\end{thm}
	\begin{pf}
		We have that the region $\mathcal{P}$
		is controlled invariant and that
		if we consider the minimizing control $u^*$ in \eqref{ineqPsi}, \mbox{for $x_0 \in \mathcal{P}$}, by integrating, we get:
		$$
		\int_0^\infty~g(x,u^*) \dd t \leq \Phi(x_0).
		$$
		This means that the infinite-horizon open loop
		optimal cost with initial condition $x_0 \in \mathcal{P}$ is upper bounded by $\Phi(x_0)$.
		Recursive feasibility  then follows from \cite{chen1998quasi}.
		
		Take any initial $x(t)=x_t$ and denote by $\bar u_t$ and $\bar x_t$
		the optimal solution on interval
		$[t,t+T]$. Consider a shifted interval
		$[\tau,\tau+T]$, with $t < \tau$ and
		take the difference
		\begin{align}
			&\Delta = \Psi_T(x(\tau))-\Psi_T(x(t)) 
			\nonumber
			\\
			&
			=\min_{u} \left \{\int_{\tau}^{\tau+T}  \hspace{-4mm}
			g(x(\sigma),u(\sigma)) \dd \sigma + \Phi(x(\tau+T))
			\right \}
			\nonumber \\
			&~~~-\min_{u} \left \{\int_{t}^{t+T} \hspace{-4mm}g(x(\sigma),u(\sigma)) \dd \sigma + \Phi(x(t+T))\right \} 
			\nonumber
			\\
			&
			=\min_{u} \left \{\int_{\tau}^{\tau+T}  \hspace{-4mm}
			g(x(\sigma),u(\sigma)) \dd \sigma + \Phi(x(\tau+T))
			\right \} 
			\nonumber \\
			&~~~-\left \{\int_{t}^{t+T} \hspace{-4mm}g(\bar x_t(\sigma),\bar u_t(\sigma)) \dd \sigma + \Phi(\bar x_t(t+T))\right \}.
		\end{align}
		Now we consider the solution $\bar u_t$ and $\bar x_t$
		in the intersection interval $[\tau,t+T]$
		in the first term, so achieving an inequality,
		and eliminate the common term
		$\int_\tau^{t+T} \hspace{-3mm}g(\bar x,\bar u) \dd\sigma$
		\begin{align}
			\nonumber
			& \Delta \leq \min_{u} \left \{\int_{\tau}^{t+T}  \hspace{-4mm}
			g( x(\sigma), u(\sigma)) \dd \sigma 
			\right . \nonumber\\
			&~~~
			+
			\left . \int_{t+T}^{\tau+T}  \hspace{-4mm}
			g(x(\sigma),u(\sigma)) \dd \sigma + \Phi(x(\tau+T))
			\right \} \nonumber
			\\
			&~~~-\left \{\int_{t}^{t+T} \hspace{-4mm}g(\bar x_t(\sigma),\bar u_t(\sigma)) \dd \sigma + \Phi(\bar x_t(t+T))\right \} \nonumber
			\\
			&=
			\min_{u}
			\left \{
			\int_{t+T}^{\tau+T}  \hspace{-4mm}
			g(x(\sigma),u(\sigma)) \dd \sigma + \Phi(x(\tau+T))
			\right \}
			\\
			&- \int_{t}^{\tau}  
			%\hspace{-4mm}
			g( \bar x_t(\sigma), \bar u_t(\sigma)) \dd \sigma 
			- \Phi(\bar x_t(t+T))
			\nonumber\\
			& =
			\underbrace{\min_{u} \int_{t+T}^{\tau+T}  \hspace{-5mm} g(x(\sigma),u(\sigma)) \dd \sigma 
				+ \Phi(x(\tau+T))
				- \Phi(\bar x_t(t+T))
			}_{\leq 0}
			\nonumber \\
			& - \int_{t}^{\tau}   
			g( \bar x_t(\sigma), \bar u_t(\sigma)) \dd \sigma
			\leq 
			- \int_{t}^{\tau}   
			g( \bar x_t(\sigma), \bar u_t(\sigma)) \dd \sigma.
			\nonumber 
		\end{align}
		The non--positivity of the term, 
		follows by considering the
		initial state $\bar x_t(t+T)
		\in \mathcal{P}$ and
		integrating \eqref{ineqPsi}
		in Assumption \ref{inequality},
		bearing in mind that $x(\sigma) \in \mathcal{P}$ if $ \sigma \in [t+T,\tau+T]$.
		
		Now we need only to reconsider $\Delta$
		divide by $\tau - t$ and take the limit
		$\tau-t \rightarrow 0$
		\begin{align}
			\nonumber
			& D^+\Psi_T(x(t),\omega(x(t))) = \limsup_{\tau-t \rightarrow 0} \frac{\Psi_T(x(\tau))-\Psi_T(x(t))}{\tau-t} \\
			\nonumber
			& = \limsup_{\tau-t \rightarrow 0} \frac{-1}{\tau-t}  \int_{t}^{\tau}   
			\hspace{-1mm}g( \bar x_t(\sigma), \bar u_t(\sigma)) \dd \sigma
			\\
			&\leq - g(x(t),\omega(x(t))).
		\end{align}
		\QEDA
	\end{pf}
	
	\begin{rem}
		We have that \eqref{gbounds}
		is not an equality as
		\eqref{HJB}.
		However, for $T$ large enough and $\mathcal{P}$
		small, there is
		no essential difference
		between the infinite and finite horizon performance, namely $\Psi (x)\approx \Psi_T(x)$,
		and
		\eqref{gbounds} becomes close to an equality and roughly, $D^+\Psi(x,\omega(x)) \approx - g(x,\omega(x))$.
		Hence, to search for an SCLF in practice, 
		we will impose the condition 
		$D^+V(x,\omega(x))  \leq -g(x,\omega(x))$ 
		rather than 
		$D^+V(x,\omega(x))  \leq D^+\Psi(x,\omega(x))$ 
		to avoid
		computing $D^+\Psi(x,\omega(x)) $.
	\end{rem}

	\section{Computation of a Sub-Control Lyapunov \\ Function}
	\label{sec:computation_sCLF}
	
	In this section, given a suitable setup, the computation of a SCLF will be cast into semi-infinite linear programming (SILP), i.e., a linear program with infinitely many constraints \cite{steentjes2020construction,julian1999parametrization,giesl2014revised,RavSan19}. Two procedures to solve such a problem will be discussed with different levels of complexity and efficiency.  
	
	One key point of the procedure
	is that we do not need to have
	the function $\Psi(x)$ to derive
	a sub-control Lyapunov function.
	We need just an algorithm to evaluate the optimal control
	$\omega(x)$ function and the given performance function $g(x,\omega(x))$
	for off--line synthesis.

	\subsection{Finding an SCLF with SILP}

	% Hereinafter, it is assumed that a control $\omega(x)$ associated with a CLF $\Psi(x)$ satisfying \eqref{ineq} is known. 
	
	Herein, under the standing assumption that the baseline control $\omega(x)$ is off-line computable, the construction of the SCLF can be carried out as follows.
	First, consider an SCLF candidate of the form:
	\begin{equation}
		V(x) = \sum_{k=1}^N\alpha_k V_k(x),
		\label{eq:para_Vx}
	\end{equation}
	where $V_k(x), \alpha_k\geq 0$ are smooth positive definite functions and the corresponding coefficients. Then, consider the following program:
	% \footnote{\textcolor{red}{ the feasible set of $\alpha$ is certainly convex, but is it bounded? }}:
	% \vspace{-0.5cm}
	\begin{subequations}
		\label{eq:ILP_SCLF}
		\begin{align}
			{\alpha}^*=&\arg\min_{\alpha\geq 0}\, \mathbf{1}^\top\alpha,\label{eq:SCLF_const}\\
			% &\alpha_k\geq 0,\\
			&\Omega({\alpha},x)\leq 0, \label{eq:SCLF_condi_ilp} \\
			&x\in \mc X \subset \R^n.\label{eq:SCLF_condi_conti}
		\end{align}
	\end{subequations}
	where ${\alpha}=[\alpha_1,...,\alpha_N]^\top $ collects the coefficients $\alpha_k$ in \eqref{eq:para_Vx}, $\mathcal{X}$ is a compact region of interest containing the origin in its interior. $\Omega(\alpha,x) :\R^N\times \R^n \rightarrow \R$ denotes the condition function:
	\begin{equation}
		\begin{aligned} 
			&\Omega({\alpha},x) \triangleq \dot V(x,\omega(x)) + g(x,\omega(x))\\
			&= \nabla V(x)(A(x) + B(x)\omega(x))+g(x,\omega(x))\\
			&=\sum_{k=1}^N\alpha_k\nabla V_k(x)(
			A(x)+B(x)\omega(x)
			) +g(x,\omega(x)).
		\end{aligned}
		\label{eq:validator}
	\end{equation} 
	The program \eqref{eq:ILP_SCLF} is a linear program with $N$ variables $\alpha_k$ and infinitely many linear constraints
	of the form \eqref{eq:SCLF_condi_ilp} with $x$ taken from $\mathcal{X}$.
	This leads to a so-called SILP which is a convex optimization problem and has been deeply investigated in the literature \cite{anderson1987linear,tanaka1988comparative,hettich1993semi,polak2012optimization}. By solving \eqref{eq:ILP_SCLF}, one can obtain the following result.
	
	\begin{prop}
		If the semi-infinite linear program \eqref{eq:ILP_SCLF} has a feasible solution $\bar \alpha = [\bar \alpha_1,...,\bar \alpha_N]^\top$, then $V(x) = \sum_{k=1}^N \bar\alpha_k V_k(x)$ is an SCLF.
	\end{prop}
	
	\begin{pf}
		The proof follows immediately by the satisfaction of \eqref{eq:SCLF_condi_ilp}--\eqref{eq:SCLF_condi_conti} and Definition \ref{def:sCLF_def}.
	\end{pf}
	
	\begin{rem}
		The choice of $\textbf{1}^\top\alpha = \sum_{i} \alpha_i$
		as cost in the linear program
		aims at finding the smallest
		value of $V(x)$ which is a CLF,
		to ensure the best performance
		(Theorem
		\ref{thm:guaran}).
		Assume that $x_0$
		is a nominal initial condition,
		representing a nominal perturbation.
		Then, as an alternative possibility, one can replace the cost function by
		$\sum_i \alpha_i V_i(x_0)$,
		to specifically improve the performance from this state. 
	\end{rem}
	
	% As a result, by solving \eqref{eq:ILP_SCLF}, one can obtain an SCLF and the online control computation is reduced from $\omega(x)$ to $\bar u(x)$ in \eqref{control}, evidently with the trade-off of optimality as mentioned in Remark \ref{rem:sub-opti}.
	% Hence, with 
	%An intuitively simple solution for \eqref{eq:ILP_SCLF} is introduced as follows.
	
	\subsection{Discretization-based solution}
	To compute a feasible solution for \eqref{eq:ILP_SCLF}, one direct approach is to sample or discretize the set $\mathcal{X}$
	into a sufficiently dense grid of points $\mathcal{X}_g \subset \mathcal X$. Then, a candidate coefficient $\alpha$ can be found by solving \eqref{eq:SCLF_const}--\eqref{eq:SCLF_condi_ilp} for $x\in \mathcal{X}_g$, making the optimization a standard Linear Program (LP):
	\begin{equation}
		\alpha^* = \underset{\alpha\geq 0}{\arg \min \, } \mathbf{1}^\top\alpha
		\text{ s.t. }\alpha\in\underset{x\in\mathcal{X}_g} \bigcap \{\alpha:\Omega(\alpha,x)\leq 0\}.
		\label{eq:discrete_constr}
	\end{equation}
	An immediate follow-up question would be how dense the grid $\mathcal{X}_g$ should be to obtain an acceptable solution within 
	a certain range of tolerance
	for the constraints in \eqref{eq:ILP_SCLF}.
	To address this concern, let us proceed with the following setup.
	Given a finite set $\mathcal{X}_g\subset \mathcal{X}$, define a granularity function:
	\begin{equation}
		\rho(x) = \underset{x_g\in\mathcal{X}_g}{\inf} \|x-x_g\|.
		\label{eq:granularity}
	\end{equation}
	It is further assumed that the time derivative of $V(x)$ driven by the control $\omega(x)$ and the corresponding penalty $g(x,\omega(x))$ is Lipschitz continuous. Namely,
	\begin{equation}
		\begin{aligned}
			&  |\Gamma(x_a) -   \Gamma(x_b) | \leq \nu \|x_a-x_b\| \\
			&  |g(x_a,\omega(x_a)) -   g(x_b,\omega(x_b))  | \leq \eta \|x_a-x_b\| ,
		\end{aligned}
		\label{eq:Lipschitz_const}
	\end{equation}
	with $x_a,x_b \in \mathcal{X}$ and for brevity, we use the notation $\Gamma(x)\triangleq \dot V(x,\omega(x))= \nabla V(x)(A(x)+B(x)\omega(x))$.
	
	Then we have the following proposition.
	
	\begin{prop}
		\label{prop:lower_bound}
		Suppose that we have a candidate $V(x)$ defined as in \eqref{eq:para_Vx} by solving \eqref{eq:discrete_constr} for a finite set $\mathcal{X}_g\subset \mathcal{X}$. Namely, $\Omega(\alpha,x_g)\leq 0~\forall x_g\in\mathcal{X}_g$, or,
		\begin{equation}
			\Gamma(x_g) \leq -g(x_g,\omega(x_g))~\forall x_g\in\mathcal{X}_g.
		\end{equation}
		Then, given $0<\varphi<1,\exists\zeta(\varphi)>0$:
		\begin{equation}
			\Gamma(x)\leq -\varphi g(x,\omega(x)),
			\label{eq:granularity_cond}    \end{equation}
		$\forall x\in \mathcal{X}\cap\{x:g(x,\omega(x))\geq \zeta(\varphi)\},$ and     $\zeta(\varphi)\triangleq(\eta+\nu)\bar \rho/(1-\varphi)$, $\bar\rho=\sup_{x\in\mathcal X}\rho(x)$.
	\end{prop}
	
	\begin{pf}
		Consider $x\in\mathcal{X}$, $x_g\in\mathcal{X}_g \cap\mathcal B_{\bar \rho}(x)$, we have:
		\begin{equation}
			\begin{aligned}
				&     \Gamma(x) =\Gamma(x_g) + \Gamma(x) -\Gamma(x_g)   \\
				&\leq - g(x_g,\omega(x_g)) + \nu \|x-x_g\|\\
				& = - g(x,\omega(x))+ g(x,\omega(x))- g(x_g,\omega(x_g))\\
				&\;\;\;\;\;\;\;\;\;\;\;\;\;\;\;\;\;\;\;\;\;\;\;\;\;\;\;\;\;\;\;\;\;\;\;\;\;\;\;\;\;\;\;\;\;\;\;\;\;\;\;\;\;+\nu \|x-x_g\| \\
				&\leq- g(x,\omega(x))+ \eta\|x-x_g\| + \nu \|x-x_g\| \\
				&\leq- g(x,\omega(x))+ (\eta  + \nu) \bar \rho \\
				% &\rho(x)(\nu+\eta )\\
				&\leq - \varphi g(x,\omega(x)),
			\end{aligned}
		\end{equation}
		where the last inequality follows from \eqref{eq:granularity_cond}. \QEDA
	\end{pf}
	This is to say, if an accuracy $0<\varphi<1$ is selected (possibly close to $1$), the inequality \eqref{eq:granularity_cond} is guaranteed continuously outside the set $\{x:g(x,\omega(x))\leq \zeta(\varphi)\}$. The volume of this level set can be shrunken by squeezing $\bar \rho$, or increasing the density of the grid $\mathcal{X}_g$.
	Proposition \ref{prop:lower_bound} hence provides a theoretical analysis on the relation between the discretization of the continuous set $\mathcal{X}$ 
	and the
	the characteristic parameters. 
	However, 
	although programmatically simple for low-dimensional systems,
	for high-dimensional systems,
	uniformly gridding the set $\mathcal{X}$ and solving \eqref{eq:discrete_constr} are not practical since the required number of points in 
	$\mathcal{X}_g$ may grow aggressively with respect to the state dimension $n$. For this reason, in the following, as opposed to a fixed grid as in \cite{johansen2000computation,do2023lp}, we establish an iterative procedure for suitably
	refining the set $\mathcal{X}_g$ based on the cutting-plane technique.
	
	\subsection{Cutting-plane-based solution}
	% The cutting plane method is an iterative algorithm for solving optimization problems. It works by successively refining the feasible region or objective function by adding linear constraints, called cuts, that eliminate non-optimal regions.
	In this part, the cutting-plane (CP) method is employed to provide an iterative procedure for grid adaptation. 
	The intuition is that of iteratively enforcing the constraint \eqref{eq:SCLF_condi_ilp} at the state where it is most violated \cite{RavSan19}. Namely, in the variable space of $\alpha$, linear constraints (a.k.a., cuts) are added correspondingly.
	In this work, the CP technique is chosen, since in the optimization problem \eqref{eq:ILP_SCLF}, there is only one continuous constraint described by \eqref{eq:SCLF_condi_ilp}--\eqref{eq:SCLF_condi_conti}. Hence, the violation can be evaluated by checking the non-positivity of the scalar function $\Omega(\alpha,x)$ in \eqref{eq:validator} over the domain $ \mathcal{X}$ for a given candidate $\alpha$ resulted from \eqref{eq:discrete_constr}, i.e., verifying:
	\begin{subequations}
		\label{eq:cert}
		\begin{align}
			&\Omega^*=\Omega(\alpha,x^*)\leq 0,  \label{eq:cert_Omg}\\        
			\text{with }&x^* = \arg \max_{x\in\mathcal{X}}\Omega(\alpha,x) . \label{eq:cert_xstr}
		\end{align}
	\end{subequations}

	% This technique successively refines the discrete grid by adding linear constraints (a.k.a., cuts) at the point where the constraint \eqref{eq:SCLF_condi_ilp} is the most violated. 
	% With this intuition, let us consider the following algorithm.
	\begin{algorithm}[ht]
		\textbf{Input:} A compact set $\mathcal{X}$, a set of basis functions $V_k(x)$, and the control law $\omega(x)$.
		
		\textbf{Output:} the SCLF $V(x)$.
		
		1:  Select an initial finite grid $\mathcal{X}_g\subset \mathcal{X}$; 
		
		2:  Solve the LP \eqref{eq:discrete_constr} with $\mathcal{X}_g$ to get $ \alpha$;
		
		3: Define the function $V(x)$ as in \eqref{eq:para_Vx};
		
		4:  Define $\Omega(\alpha,x)$ as in \eqref{eq:validator};
		
		5:  Compute $x^*$ and $\Omega^*$ as in \eqref{eq:cert};
		
		6:  \textbf{if} $\Omega^*>0$
		
		7:  \hspace{0.7cm} Append $\{x^*\}$ to the the grid points $\mathcal{X}_g$;
		
		8:  \hspace{0.7cm} \textbf{go to} step 2;
		
		9:  \textbf{else}
		
		10:  \hspace{0.7cm}\textbf{return} $V(x)$;
		
		\caption{SCLF construction with CP method}
		\label{algo:CP}
	\end{algorithm}
	% circumvents circumvents circumvents circumvents circumvents circumvents 

\begin{figure}[htbp]
		\centering
		\resizebox{0.475\textwidth}{!}{\input{pics/tikz/algo_cutplane_automatica}}
		\caption{Procedure for the SCLF construction based on cutting-plane technique.}
		\label{fig:algo_cutplane_automatica}
	\end{figure}
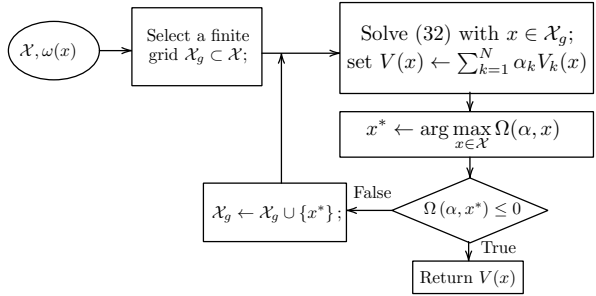

	If the certificate \eqref{eq:cert_Omg} does not hold, $x^*$ in found \eqref{eq:cert_xstr} is the point where the constraint $\Omega(\alpha,x)\leq 0$ \eqref{eq:SCLF_condi_ilp} is violated the most. This point will then be appended to the current grid $\mathcal{X}_g$ and the procedure will be reinitialized with such an updated grid.
	Concisely, Algorithm \ref{algo:CP} summarizes the main steps of the SCLF construction based on CP technique, an illustrative flow chart is also provided in Fig. \ref{fig:algo_cutplane_automatica}.

	% \begin{RevisedNote}
		\begin{rem} \label{rem:nonconvex}
			Solving \eqref{eq:cert_xstr} is not straightforward because the optimization problem is almost never convex. However, note that one has the non-convexity only in the state space $x$ when solving \eqref{eq:cert_xstr}. 
			Therefore, even though non-trivial, with proper solvers, the algorithm can be applied for practical scenarios (see later in Section \ref{sec:numericalExp}).
			Meanwhile, in the coefficient space $\alpha$ of high dimension, the problem is a standard LP in \eqref{eq:discrete_constr}.
		\end{rem}
		% \end{RevisedNote}

	In the following part, to proceed with the construction of the SCLF for the numerical examples, we particularize the choice of the  $\omega(x)$ and the basis functions $V_k(x)$.
	% As also illustrated in Fig. \ref{fig:algo_cutplane_automatica}, the key component for the procedure is the verification of whether the scalar function $\Omega(\alpha,x)$

	\subsection{Discussion}

	\textit{i) The ideal control $\omega(x)$}:
	
	To obtain a control law of optimal performance, among practical control strategies, quasi-infinite-horizon MPC \cite{chen1998quasi} will be chosen as the software for generating stable closed-loop trajectories due to its standardized synthesis procedure. 
	
	For numerical implementation of the software, the constraint \eqref{eq:dyna_qih} is approximated via the Euler discretization \cite{blanchini2003suboptimal} with a sampling parameter $\tau_s>0$.
	
	The next fundamental component will be the set of basis functions parameterizing the SCLF. In order to provide a rationale for our selection of such a set for the numerical examples, a brief review will be presented subsequently.

	\textit{ii) The basis functions $V_k(x)$ choice}
	
	In the literature, the parameterization of the form \eqref{eq:para_Vx} to find a CLF has been intensively studied. Hence, there are several options for the choice of the basis functions $V_k(x)$.
	Generally, this is often a trade-off between their flexibility and computational complexity in both offline synthesis and online deployment stages. 
	For example, in the context of artificially generated CLF, one customary choice is a set of piecewise-affine (PWA) functions, making the synthesis abridge to an LP, accompanied by generalized analytical and practical guarantees \cite{rubagotti2016lyapunov,giesl2014revised,steentjes2020construction}. 
	Yet, if the processing power is limited, one encounters the problem of dimensionality again while dividing the state space into partitions (e.g., triangulations) during the offline phase. Meanwhile, the online phase faces the so-called \textit{point location problem} when PWA functions are employed in the form of look-up tables. In the context of stability analysis for autonomous systems, a more complete set was used in \cite{johansen2000computation} in the form of a Gaussian basis. This can be thought of as an interpolation of smooth quadratic functions, with the weight matrices being the decision variables, resulting in, again, an SILP. 
	In this manner, the problem of imposing continuity on the function was also sidestepped. 
	However, with this interpolation, the basis is again defined over a set of discrete knots, complicating the online applications with respect to the system's size.
	Hence, depending on the computational power at hand, different choices of basis can be combined to acquire a satisfactory characterization.
	With these analyses, in view of maintaining smoothness, avoiding discretization-based interpolation, and taking advantage of the results for linear systems in \cite{blanchini1999new,do2023lp}, we examine the set of basis functions composed as follows.
	
	Let us first include in the basis a quadratic function:
	\begin{equation}
		V_1(x) = x^\top P x, 
		\label{eq:quad}
	\end{equation}
	with the weight $P\succ 0$ computed from the algebraic Riccati equation for the linearly approximated dynamics of \eqref{sys} around the origin. In this manner, we ensure that, around a neighborhood of the origin, we obtain some local optimality with the resulting function.
	
\begin{figure}[htbp]
		\centering
		\includegraphics[width=0.475\textwidth]{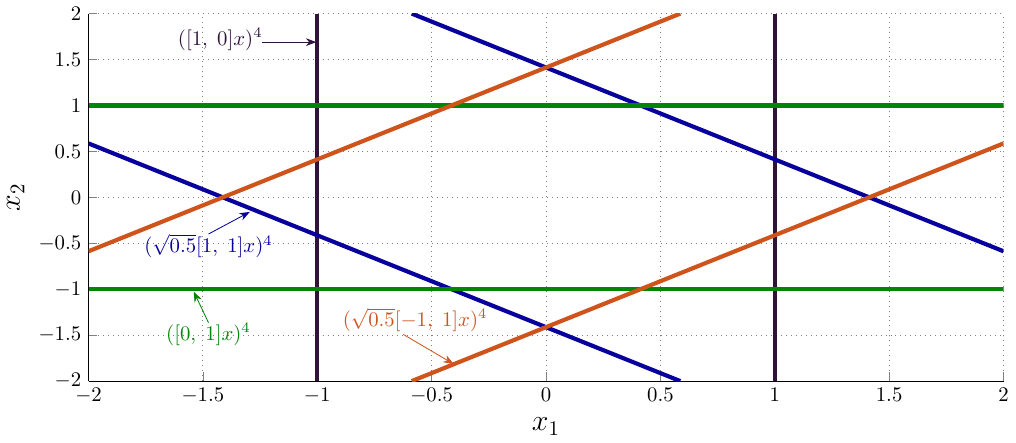}
		\caption{Unit level set of some basis functions of the form \eqref{eq:poly2p} and a linear combination of them (red), with $x\in\R^2, p = 2$.}
		\label{fig:illusbasis}
	\end{figure}

	Next, the set of basis functions will be populated with $2p$-norm-like polynomial \cite{blanchini1999new}:
	\begin{equation}
		V_k(x) = (a_k^\top x)^{2p},
		\label{eq:poly2p}
	\end{equation}
	where the function is characterized by the choice of the vector $a_k\in\R^n$. With this setting, by simply collecting $a_k$ with different directions and lengths, we can effectively expand the set of basis functions to achieve more representational capacity.
	An illustration in $\R^2$ is given in Fig. \ref{fig:illusbasis}.
	
	With the characterization of $V_k(x)$ in \eqref{eq:quad}--\eqref{eq:poly2p}, the idea is to increase the completeness of the basis with a possibly large number of basis (i.e., $N$ in \eqref{eq:para_Vx}). 
	This is the expense of fixing a priori the structure (shape and value) of the basis, as in \eqref{eq:poly2p}. Yet, it will be shown later in the examples that, eventually, after the generation of the SCLF, there exist many basis functions associated with insignificantly small coefficients.
	Hence, in practice, those components of the parameterization can be discarded. In the numerical examples, 
	a threshold $\varepsilon_N \ll 1$ will be defined so that only the coefficients $\alpha_k \geq \varepsilon_N$ are kept. The number of those remaining functions with will be denoted as $\underline N$.
	These functions will then be retained for the online control via \eqref{control}.
	
	So far, all the ingredients for the control synthesis with an SCLF are ready. Next, numerical examples will be provided with both simulation and experimental results, showing that the approach can be applied to non-trivial cases.
	\section{Numerical examples and experimental\\validation}
	\label{sec:numericalExp}
	
	Herein, three systems will be examined to illustrate and showcase the effectiveness of the proposed machinery. First, the dynamics investigated will be recapitulated, and then the numerical results will be reported. Note that in the next part, without further explanation, the notation $x,u$ will be abused to denote the state and input vector of the system under discussion.

	\subsection{System description}
	
	Let us proceed by briefly summarizing the three control problems employed, as follows.

	\textit{i) Bioreactor process (Bioreactor)}
	
	For the sake of illustration, we first examine a two-dimensional system of a bioreactor process governed by the following nonlinear dynamics \cite{brengel1989multistep,ramaswamy2005control}:
	\begin{equation}
		\begin{aligned}
			& \dot x_1  = -(C_{1S} + x_1)(1+u) + \\
			&~~~~ D_{aS}(1-x_2-C_{2S})e^{(C_{2S}+x_2)/\gamma}(C_{1S} + x_1)\\
			& \dot x_2 =-(C_{2S} + x_2)(1+u) + \\
			&  D_{aS}(1-x_2-C_{2S})e^{(C_{2S}+x_2)/\gamma}\frac{(1+\beta)(C_{1S}+x_1)}{1+\beta-x_2-C_{2S}}.
		\end{aligned}
		\label{eq:bio_sys}
	\end{equation}
	
	with the parameters $\gamma = 0.4,\beta = 0.02,  D_{aS} = 1.2$, $ C_{1S}=0.0938,C_{2S}= 0.9155.$ $x_1$ and $x_2$ denote the deviation from the equilibrium dedimensionalized states, while the dimensionless flow rate $u$ is the system's input and its amplitude is constrained as:
	\begin{equation}
		|u| \leq u_{max } =1.
		\label{eq:bio_cons}
	\end{equation}
	The control objective for system \eqref{eq:bio_sys} is to stabilize $x_1,x_2$ towards the origin. 
	The domain of interest is set as $\mathcal{X}=\{x: |x|\leq 0.1\}$.
	With this model, it is important to note that the state, the input and even the time clock have been normalized. However, for consistency, the same notations for the variables and their derivatives will be used.
	For the technical analysis of the model construction, we would like to send the readers to \cite{brengel1989multistep,ramaswamy2005control} and the references therein.

	% Note that in this model, although denoted with the same notation for the time variable $t$ and the time derivative the time variable is normalized
	
	% % Table generated by Excel2LaTeX from sheet 'Sheet1'
	% \begin{table}[htbp]
		%   \centering
		%   \caption{Specifications and total build time (BT) for the three systems}
		%     \begin{tabular}{|c|c|c|c|c|c|}
			%     \hline
			%     Model & nonlinear & n    & m    & N    & \multicolumn{1}{p{3em}|}{Total\newline{}BT (h)} \\
			%     \hline
			%     Bioreactor  &  \xmark   &  2    &  1    &  1303    & 3.144 \\
			%     \hline
			%     VTOL &   \xmark  &   6   &   2   &     1627 &  22.59 \\
			%     \hline
			%     Drone &   \cmark    &   6   &  3    &   2561   &    80.93\\
			%     \hline
			%     \end{tabular}%
		%   \label{tab:3sys}%
		% \end{table}%
	
	\textit{ii) Planar vertical take-off and landing aircraft (VTOL)}
	
\begin{figure}[hpbt]
		\centering
		\includegraphics[width=0.4\textwidth]{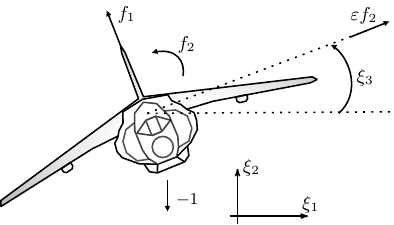}
		\caption{The planar VTOL model}
		\label{fig:vtol_draw}
	\end{figure}

	Next, to showcase the effectiveness of the technique over a non-trivial size system, we examine the normalized model of a vertical take-off and landing (VTOL) aircraft driven by the following dynamics \cite{hauser1992nonlinear,martin1996different}:
	\begin{equation}
		\begin{aligned}
			\ddot \xi_1 &= -f_1\sin \xi_3 + f_2\varepsilon  \cos \xi_3, \\
			\ddot \xi_2 &= f_1\cos \xi_3 + f_2\varepsilon  \sin \xi_3  -1 ,\\ 
			\ddot \xi_3 & = f_2,
		\end{aligned}
		\label{eq:vtolModel}
	\end{equation}
	where in \eqref{eq:vtolModel}, $\xi_1,\xi_2$ and $\xi_3$ denote the 2-D positions of the aircraft's center of mass and its roll angle, respectively.  $\varepsilon=0.2$ is the system's parameter, and ``$-1$" represents the normalized gravitational acceleration. Finally, $f=[f_1,\; f_2]^\top$ collects the thrust and rolling moment (see Fig. \ref{fig:vtol_draw}). The control objective here is to stabilize the system at the equilibrium point where:
	\begin{equation}
		\xi_i=0,\dot \xi_i = 0, i\in\{1,2,3\}, \text{ and }f=f_e=[1,\;0]^\top.
		\label{eq:VTOL_equi}
	\end{equation}

	To verify the SCLF found with the explicit control \eqref{eq:closedForm_control}, the constraint on the input $f$ is intentionally chosen as:
	\begin{equation}
		[-2,\;-2]^\top  \leq f \leq [4,\;2]^\top.
	\end{equation}
	In this way, the error dynamics can be rewritten in the control affine form as:
	\begin{equation}
		\begin{aligned}
			&  \dot x_1 =x_2, \dot x_2 = -(u_1+1)\sin x_5+ u_2\varepsilon  \cos x_5 ,\\
			&\dot x_3 =x_4, \dot x_4 = (u_1+1)\cos x_5 + u_2\varepsilon  \sin x_5  -1 ,\\
			&\dot x_5 = x_6,\dot x_6 = u_2. 
		\end{aligned}
		\label{eq:vtol_error_dyna}
	\end{equation}
	with $x_{2i-1} = \xi_{i}, i\in\{1,2,3\}$,
	while the input $u=f-f_e$ is restricted in the form of \eqref{eq:box_U}:
	\begin{equation}
		|u|\leq u_{max} = [3,\;2]^\top.
		\label{eq:vtol_error_cons}
	\end{equation}
	% [1 0.5 1 0.5 pi/3 pi/6];
	The investigated state space for this system is set as: 
	\begin{equation}
		\begin{aligned}
			& \mathcal{X} = \Big\{x: |x_1|,|x_3|\leq 1, |x_2|,|x_4| \leq 0.5, \\ 
			& \quad\quad\quad |x_5|\leq \pi/3 , x_6\leq \pi/6 \Big \}.
		\end{aligned}
	\end{equation}

	\begin{table*}[th]
		\caption{Numerical parameters used in the SCLF synthesis and online control}
		\centering
		\begin{tabular}{|c|c|c|c|c|c|c|c|}
			\hline
			& System \& constraint &$Q$ & $R$ & $\tau_s$ (s) & $n_g$  &$T^{\text{off}}$ (s) & $T^{\text{on}}$ (s)    \\ \hline
			Bioreactor      &  \eqref{eq:bio_sys} and \eqref{eq:bio_cons}   & diag$(5,5)$  & 1 &  0.05 & 20 &17.5  & 0.5 \\
			VTOL      & \eqref{eq:vtol_error_dyna} and \eqref{eq:vtol_error_cons}& diag$([5,1,5,1,10,1])$  & diag$(4,4)$ &  0.05 & 1500 &15 & 3.25 \\
			Drone      & \eqref{eq:ThreeDIs} and \eqref{eq:Vctilde}& diag$([50,50,50,5,5,5])$  & diag$(10,10)$  & 0.1 & 1500 & 12 & 3.5\\\hline
		\end{tabular}
		\label{tab:num_spec}
	\end{table*}

	\textit{iii) Quadcopter position control (Drone)
	}
	
	The final example will be the experimental validation for the position control (a.k.a., outer-loop control) of a nano-drone \cite{do2023lp}. The dynamics can be given as follows:
	\begin{equation}
		\begin{aligned}
			&\ddot x_1 = F(\cos\phi \sin\theta\cos\psi  + \sin\phi\sin\psi), \\
			& \ddot x_2 = F(\cos\phi \sin\theta\sin\psi  - \sin\phi\cos\psi),\\
			&\ddot x_3 = -\underline{\mathsf{g}} + F\cos\phi \cos\theta,
		\end{aligned}
		\label{eq:DroneDyna}
	\end{equation}
	where 
	$x_i,i={1,2,3}$ are the drone's 3-D positions, 
	$\underline{\mathsf{g}}$ is the gravitational acceleration. $(\phi,\theta,\psi)$ denote the roll, pitch, and yaw angles, respectively. $F\geq 0$ is the normalized thrust provided by the propellers.
	In this setting, $\psi$ is assumed to be known by measurement, while $F,\phi,\theta$ are the manipulated variables and constrained as: 
	\begin{equation}
		0\leq F\leq \bar F, |\phi |\leq \bar \epsilon,|\theta |\leq \bar \epsilon
		,
		\label{eq:constr_original_drone}
	\end{equation}
	with $\bar F=1.45\underline{\mathsf{g}}$ and $ \bar \epsilon=0.1745$ (rad) being the bounds of the thrust and the angles.
	These signals are then provided to the built-in inner-loop which controls the propellers with a proper conversion. Details for this hierarchical setting can be found in \cite{huang2021closed}.
	Then, the objective is to stabilize the drone at the equilibrium point, where:
	\begin{equation}
		% \begin{aligned}
			x_i=0,\dot x_i = 0, F = \underline{\mathsf{g}}, \phi = \theta =0.
			% \end{aligned}
	\end{equation}
	For that purpose, it was known that \eqref{eq:DroneDyna} can be exactly linearized via the input transformation:
	\begin{subequations}
		\begin{align}
			F&=\sqrt{u_1^2+u_2^2+(u_3+\underline{\mathsf{g}})^2},\label{eq:linearization_a} \\
			\phi&=\arcsin{\left({(u_1\sin{\psi}-u_2\cos{\psi})}/{F}\right)},\label{eq:linearization_b} \\
			\theta&=\arctan{\left({(u_1\cos{\psi}+u_2\sin{\psi})}/{(u_3+\underline{\mathsf{g}})}\right)}.\label{eq:linearization_c}
		\end{align}
		\label{eq:linearization}
	\end{subequations}

	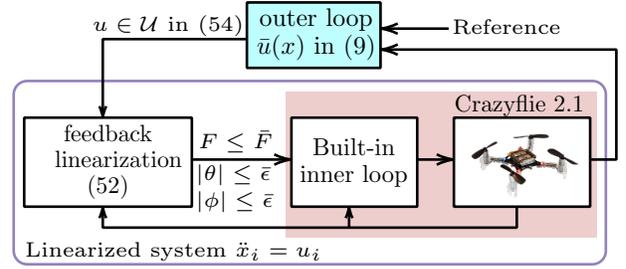
\begin{figure}[htbp]
		\centering
		\resizebox{0.47\textwidth}{!}{\input{pics/tikz/droneHier}}
		\caption{Hierarchical control scheme of the experiments.}
		\label{fig:droneHier}
	\end{figure}

	Indeed, replacing \eqref{eq:linearization} in \eqref{eq:DroneDyna} yields:
	\begin{equation}
		\dot x_i = x_{i+3}, \dot x_{i+3} = u_i, i \in \{1,2,3\},
		\label{eq:ThreeDIs}
	\end{equation}
	and the constraint \eqref{eq:constr_original_drone} propagated through the mapping \eqref{eq:linearization} can be under-approximated by the polytope \cite{do2023lp}:
	\be 
	\mathcal{U}= \mathrm{co}
	\left\{
	\begin{aligned}
		\relax    [0,0,-\underline{\mathsf{g}}]^\top, [r^\star\cos\mu, r^\star\sin\mu,u_3^\star]^\top&\\
		[r\cos\mu,r\sin\mu,\sqrt{\bar F^2-r^2}-\underline{\mathsf{g}}]^\top&
	\end{aligned}
	\right\},
	\label{eq:Vctilde}
	\ee 
	with 
	$r^\star=\bar F\sin\bar \epsilon$, $u_3^\star=\bar F\cos\bar \epsilon-\underline{\mathsf{g}}$
	and for some integers $S_1,S_2>2$,  $\mu\in \mathcal{N}(S_1,0,2\pi),r\in \mathcal{N}(S_2,0,r^\star)$. For this work, we chose $S_1=8,S_2=2.$

	In brief, the control problem now involves the stabilization of system \eqref{eq:ThreeDIs} subject to $u=[u_1,u_2,u_3]^\top\in\mathcal{U}$ as in \eqref{eq:Vctilde}. 
	The studied domain is chosen as: 
	\begin{equation}
		\mathcal{X}=\{x: |x|\leq 1.5\}.
	\end{equation}
	The complete scheme for this application is depicted in Fig. \ref{fig:droneHier}. Later, for this system, the experiments will be carried out on the Crazyflie 2.1 nano-drone in an indoor environment. Feedback signals will be estimated with the Qualisys motion capture system. For brevity, we would like to send the readers to our previous experiments in \cite{do2023lp} for further details on the experimental platform.
	In short, the three systems and their constraints are summarized in the first two columns of Table \ref{tab:num_spec}.

	% (the dynamics \eqref{eq:vtolModel} translated to the equilibrium point in \eqref{eq:VTOL_equi}) will have a box-type constraint as in \eqref{eq:box_U}, highlighting the effectiveness of the explicit minimizer \eqref{eq:closedForm_control}.
	
	\subsection{Numerical results and discussion}
	
	In this part, the numerical results for the offline synthesis of the SCLF will be reported first. Subsequently, the applicability of the constructed SCLF will be demonstrated via simulations and experiments in comparison with the quasi-infinite-horizon MPC.

	\textit{i) Offline synthesis for SCLFs}
	
	% As illustrated in Fig. \ref{fig:overallScheme}, t
	The offline construction will be carried out with the control software $\omega(x)$ calculated with the routine \eqref{eq:qihMPC}. For this setup, the cost function $g(x,u)$ will be chosen in the quadratic form as in \eqref{eq:quad_cost}. The corresponding terminal cost and constraints ($\Phi(x)$ and $\mathcal{P}$) therein will be uniquely determined as the quadratic function from the linear quadratic regulator and its ellipsoidal level set computed for the approximated dynamics at the origin \cite{chen1998quasi}. Furthermore, as this is an offline effort, the prediction horizon in \eqref{eq:qihMPC} is preferably chosen to be relatively large and denoted as  $T = T^{\text{off}}$. Note that the superscript ``off" is to distinguish the value from the prediction horizon used later for online comparison. To start Algorithm \ref{algo:CP}, the initial grid was chosen as a set of $n_g$ random points in $\mathcal{X}$.
	All the aforementioned values are reported in Table \ref{tab:num_spec}. 
	All the receding horizon routines and quadratic programs will be solved with IPOPT software \cite{wachter2006implementation} via CasADi interface \cite{andersson2019casadi}. To evaluate the certificate \eqref{eq:cert}, \texttt{fmincon} solver of MATLAB 2021b will be used with the sequential quadratic programming (SQP) approach.

	\begin{table}[htbp]
		\centering
		\caption{Specifications and total build time (BT) for the three systems}
		\begin{tabular}{|c|c|c|c|c|c|c|}
			\hline
			Model  & $n$    &$ m$    & \multicolumn{1}{p{3em}|}{$\;\;\;N$\newline{}in \eqref{eq:para_Vx}}  & $\varepsilon_N$ & $\underline N$& \multicolumn{1}{p{3em}|}{Total\newline{}BT (h)}\\
			\hline
			Bioreactor     &  2    &  1    &  1303  & $5\mathrm{e}{-7}$ & 6 & 3.144\\
			
			VTOL   &   6   &   2    &     1627 &  $1\mathrm{e}{-5}$ & 29 &22.59 \\
			
			Drone     &   6   &  3    &   2561   & $6\mathrm{e}{-6}$  & 38 & 80.93\\
			\hline
		\end{tabular}%
		\label{tab:3sys}%
	\end{table}%
	% \begin{figure*}[htp]
		%     \centering
		%     \includegraphics[width=0.85\linewidth]{pics/img/loss3sys.pdf}
		%     \caption{Performance loss index for the three systems (defined in \eqref{eq:performanceLoss}) caused by the employment of the SCLF-based controller \eqref{control} over the optimal control \eqref{eq:qihMPC}. }
		%     \label{fig:loss3sys}
		% \end{figure*}

\begin{figure}[htbp]
		\centering
		\resizebox{0.47\textwidth}{!}{\input{pics/tikz/SCLF_3sys}}
		
		\caption{Largest violation found and build time (BT) for Bioreactor, VTOL and Drone model (top to bottom, respectively).}
		\label{fig:SCLF_3sys}
	\end{figure}
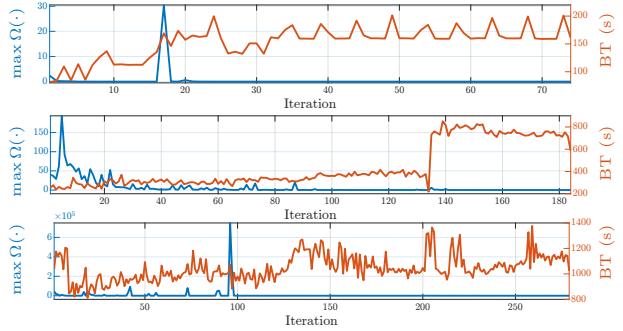

	Regarding the results in Table \ref{tab:3sys}, we recall that $N$ denotes the number of basis functions employed in the optimization \eqref{eq:ILP_SCLF}. $\varepsilon_N$ is the chosen threshold below which all the coefficients $\alpha_k$ found will be discarded, leaving $\underline N$ functions parameterizing $V(x)$ as in \eqref{eq:para_Vx}.
	The total build time (BT) used for running Algorithm \ref{algo:CP} is also reported therein. 
	Fig. \ref{fig:SCLF_3sys} the values of $\Omega^*$ in \eqref{eq:cert_Omg} found (blue) and the corresponding build time (orange) to find them along the iteration steps of the algorithm.

	\begin{figure}[ht]
		\centering
		\includegraphics[width=0.425\textwidth]{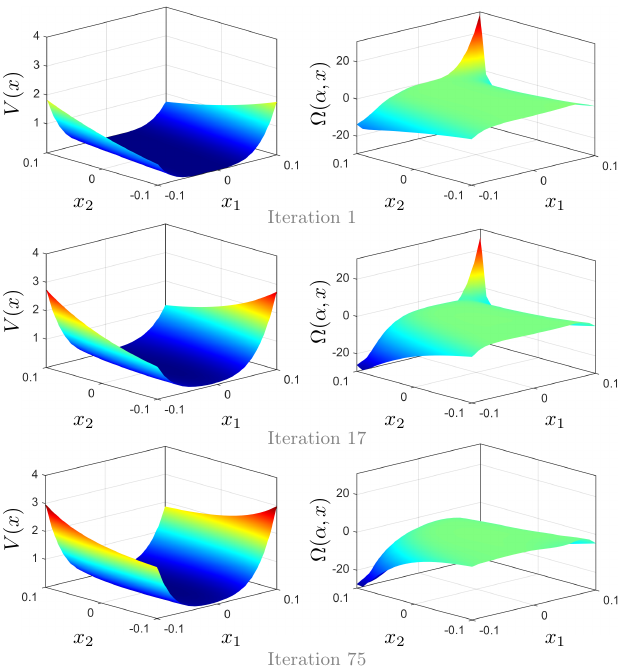}
		\caption{Snapshots of the SCLF candidates (left) and the verifier $\Omega(\alpha,{x})$ as in \eqref{eq:validator} (right) during the execution of Algorithm \ref{algo:CP} for the bioreactor model \eqref{eq:bio_sys}.}
		\label{fig:bioreact_sCLF}
	\end{figure}
	Expectedly, the offline construction time grows with respect to the number of states and inputs. This can be explained by the search for $x^*$ in \eqref{eq:cert_xstr} which requires a continuous evaluation of $\omega(x)$ within the SQP solver employed.
	It is also noticeable in Table \ref{tab:3sys} that, although using a large number of basis functions to find $V(x)$, the number of meaningful coefficients remaining is relatively small. This is, indeed, a positive sign, since later on, during the online implementation of \eqref{control}, the calculation of $\nabla V(x)$ will require less computational power.
	Regarding the effectiveness of Algorithm \ref{algo:CP}, it can be seen from Fig. \ref{fig:SCLF_3sys} and Table \ref{tab:num_spec} that the total number of points\footnote{the total of the number of points in the initial grid $n_g$ and the number of iterations.} in state space used to enforce the constraint \eqref{eq:SCLF_condi_ilp} is relatively small, in comparison with, for instance, a loose grid of 5 samples ($5^n$ points) in each axis.

	Particularly for the Bioreactor model \eqref{eq:bio_sys}, the numerical result from Algorithm \ref{algo:CP} is:
	\begin{equation}
		\begin{aligned}
			V(x) =  1.957 x^\top \begin{bsmallmatrix}
				16.973 &  -0.147 \\
				-0.147  &  1.747     
			\end{bsmallmatrix} x + 19.003([5,\, 0]x)^4 & \\
			+ 0.229([-2.823,\,4.127]x)^4 +1.074([-3.141,\, 3.890]x)^4 & \\
			+ 0.833([-3.687,\, 3.377]x)^4    + 9.003([-4.950,\, 0.706]x)^4.&
		\end{aligned}
	\end{equation}
	% 1.9574   19.0031    0.2289    1.0739    0.8333    9.0027

	With this function, Fig. \ref{fig:SCLF_3sys} and \ref{fig:bioreact_sCLF} also illustrate that the non-positivity of the surface $\Omega(\alpha,x)$ (i.e., the satisfaction of \eqref{eq:SCLF_condi_ilp}) is gradually enforced as Algorithm \ref{algo:CP} goes through its steps, especially after the 17th iteration.

	\textit{ii) Online control with the generated SCLFs}
	
	After achieving the function $V(x)$ from the offline stage, let us proceed by verifying the proposed control via simulations and experiments. The first performance indicator we would like to use is Theorem \ref{thm:guaran}, or particularly the property in \eqref{eq:dev_V_minus_g}.
	With this inequality, we can show that the closed-loop system is asymptotically stable and verify the performance analysis in the theorem. For this reason, let us denote:
	\be 
	\begin{aligned}
		&\Theta(x) \triangleq \dot V(x, \bar u(x))  + g(x,\bar u(x))\\ 
		&=\nabla V(x)(A(x)+B(x)\bar u(x)) + g(x,\bar u(x)),
	\end{aligned}
	\label{eq:thetaSum}
	\ee 
	the sum of the time derivative of $V(x)$ along the closed-loop trajectory driven by $u=\bar u(x)$ as in \eqref{control} and the corresponding cost function $g(x,u)$. Then, the system's convergence can be shown via the non-positivity of $\Theta(x) $.
	
	\begin{figure}[htb]
		\centering
		\resizebox{0.45\textwidth}{!}{\input{pics/tikz/bioreact_sim}}
		\caption{Closed-loop trajectories (blue) and the level set of the SCLF (dotted black) of the bioreactor model.}
		\label{fig:bioreact_sim_label}
	\end{figure}
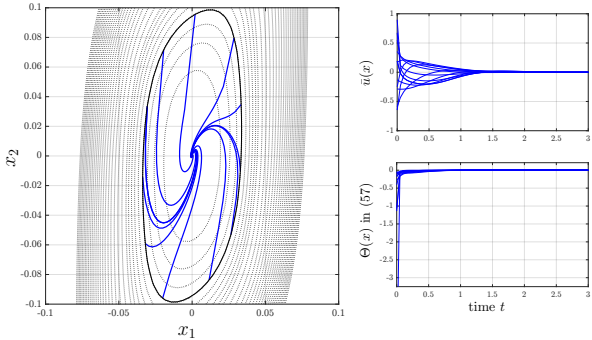
	
	For the digital implementation, taking advantage of the input constraint's shape, we employed the explicit control law in Proposition \ref{prop:sat_control_R}
	for the Bioreactor and VTOL models.
	With a non-standard input constraint as in \eqref{eq:Vctilde} for the Drone, during the experiments, the QP \eqref{control} was solved directly via IPOPT solver.

	% \clearpage

	% \subsection{Vertical take-off and landing aircraft}
	% A normalized model of a vertical take-off and landing (VTOL) aircraft 
	% \cite{hauser1992nonlinear,martin1996different}:
	
	% \begin{equation}
		%     \begin{aligned}
			%         \ddot x_1 &= -u_1\sin x_3 + u_2\varepsilon  \cos x_3 \\
			%  \ddot x_2 &= u_1\cos x_3 + u_2\varepsilon  \sin x_3  -1 \\ 
			%  \ddot x_3 & = u_2.
			%     \end{aligned}
		% \end{equation}
	
	% \begin{figure}[htbp]
		%     \centering
		%     \includegraphics[width=0.375\textwidth]{pics/pdf/vtol_draw.pdf}
		%     \caption{The planar VTOL model}
		%     \label{fig:vtol_draw}
		% \end{figure}

	\begin{figure}
		\centering
		\resizebox{0.475\textwidth}{!}{\input{pics/tikz/vtol_sim_1col_clean}}
		\caption{Planar VTOL stabilization with input constraints.}
		\label{fig:vtol_sim}
	\end{figure}
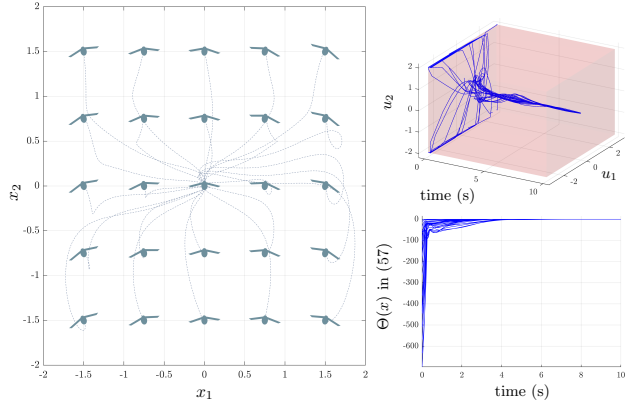
	
	To highlight the computational advantage of the proposed control, for the same initial states, the routine \eqref{eq:qihMPC} is considered for online control with the same choice of cost and constraints as in the offline synthesis part. The difference is only the horizon size, which is chosen as the minimum value of $T$
	offering a feasible solution for the tested initial states. This value is denoted as $T^{\text{on}}$ and given in Table \ref{tab:num_spec}.
	Then, the root-mean-square error (RMSE) will be computed as in \eqref{eq:rmse} to evaluate how the controls perform:
	\be 
	\text{RMSE} = \sqrt{\dfrac{1}{N_s}
		\sum_{k=1}^{N_s}\left\|x(kt_s)\right\|^2}
	\label{eq:rmse}.
	\ee 
	
\begin{figure}[hbtp]
		\centering
		\resizebox{0.475\textwidth}{!}{\input{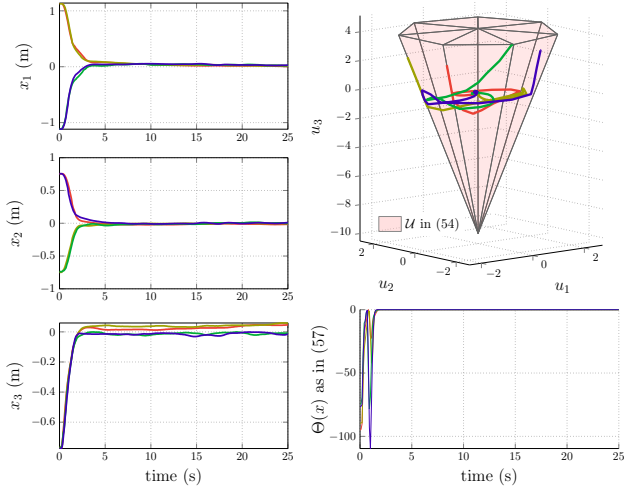}}
		\caption{Experiments for the quadcopter position tracking with the generated SCLF.}
		\label{fig:exp_drone}
	\end{figure}

	Fig. \ref{fig:bioreact_sim_label}, \ref{fig:vtol_sim} and \ref{fig:exp_drone} present the closed-loop trajectories with the stability indicator $\Theta(x)$ as in \eqref{eq:thetaSum}.
	Therein, it can be seen that with the SCLFs artificially generated before, the constrained systems are successfully stabilized, and the guarantee of $\Theta(x)\leq 0$ is also confirmed.
	
	\begin{table}[bp]
		\centering
		\caption{Comparison between SCLF-induced controller and MPC}
		\begin{tabular}{|c|c|c|c|c|}
			\hline
			\multirow{2}[5]{*}{Model} & \multicolumn{2}{c|}{SCLF} & \multicolumn{2}{c|}{MPC} \\
			\cline{2-5}         & CT (ms)  & \multicolumn{1}{p{3em}|}{Avg.\newline{}RMSE} & CT (ms)  & \multicolumn{1}{p{3em}|}{Avg.\newline{}RMSE} \\
			\hline
			Bioreactor  &    0.031  &   0.106   &    4.662   &    0.088\\
			
			VTOL &  0.207    &      0.706  &     22.174   &      0.6303
			\\
			
			Drone &   6.51   &  0.393  &   32.66      &   0.396\\
			\hline
		\end{tabular}%
		\label{tab:compare}%
	\end{table}%

	\begin{figure}[hbt]
		\centering
		\includegraphics[width=0.925\linewidth]{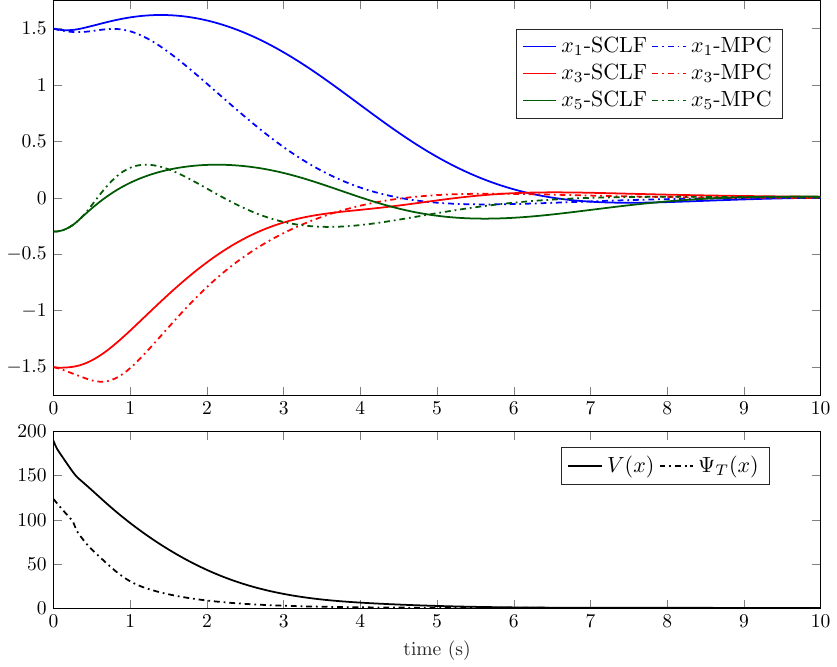}
		\caption{Suboptimal performance of the SCLF controller w.r.t. MPC in \eqref{eq:qihMPC} for the VTOL model.}
		\label{fig:comparePsiV}
	\end{figure}
	
	Regarding the performance, the trajectory of the VTOL from  $x_0=[1.50,   -0.19,  -1.50,   -0.17,  -0.30,   -0.02]^\top$ is reported in Fig. \ref{fig:comparePsiV} for both the SCLF control \eqref{control} and MPC \eqref{eq:qihMPC}. The suboptimal transient behavior of the SCLF controller is evident through a larger rise time for the states. Additionally, interpreting the cost-to-go function's upper bound, the SCLF $V(x)$ consistently exhibits a larger value compared to $\Psi_T(x)$ \eqref{eq:qihMPC}. To quantify the performance loss, we approximate the achievable cost of the infinite horizon formula \eqref{opti_infty} with Euler discretization of 1ms and 50s prediction horizon, yielding $ \Psi_\infty(x_0)\approx  62.96 .$
	Meanwhile,
	its upper bounds are given by the finite-horizon cost and the SCLF with
	$  \Psi_T(x_0) = 123.55$ and $ V(x_0) = 189.62$, gauging the trade-off to reduce implementation complexity. Indeed, although
	the proposed control deduced from the SCLFs showed slightly larger tracking error compared to the MPC law \eqref{eq:qihMPC}, its computational simplicity is a significant advantage. This simplicity is not only apparent with the explicit solution for the Bioreactor and VTOL, but also with the implicit law \eqref{control} for the Drone system, which requires significantly less computation time (CT) (See Table \ref{tab:compare}).

	% \begin{figure}[hbt]
		%     \centering
		%     \includegraphics[width=0.925\linewidth]{pics/pdf/comparePsiV.pdf}
		%     \caption{Suboptimal performance of the SCLF controller w.r.t MPC}
		%     \label{fig:comparePsiV}
		% \end{figure}

	Finally, with the experiments on the Drone model, although satisfactory tracking is shown, one practical shortcoming of the proposed control \eqref{control} is revealed with the steady state error observed in Fig. \ref{fig:exp_drone}. 
	More specifically, there has not been any robustification measure taken into account
	in both offline and online phases. One possible direction is to incorporate the SCLF with an observer, compensating for the disturbance and uncertainty online.

	\section{Conclusion and outlook}
	\label{sec:Conclusion}
	This paper addresses a control problem for which there exists a controller of high performance but, at the same time, requiring excessive computational overhead.
	Our alternative solution centers around exploiting such an ideal control law to generate a stability certificate with suboptimal performance, called the Sub-Control Lyapunov Function (SCLF).
	With the SCLF, a more computationally low-cost control is derived, accompanied by an ancillary analysis of stability and suboptimality. Finally, a computational method is also provided to virtually generate the function based on the cutting-plane technique.
	The results are then successfully validated via both simulations and experiments on systems of non-trivial size and non-linearity. The future direction will focus on the robustification of the scheme, in not only the offline synthesis but also the online implementation, in conjunction with disturbance countermeasures.

	% \clearpage
	\begin{ack}                               % Place acknowledgements
		This work is funded 
	\end{ack}

	% \appendix
	% \section{Numerical parameters employed in the synthesis and online control}    % Each appendix must have a short title.
	% \bibliographystyle{plain}        
	% \bibliography{Bibs/bib}     

\input{Arxiv.bbl}
\end{document}

%% file: pics/tikz/algo_cutplane_automatica.tex
\tikzset{every picture/.style={line width=0.75pt}} %set default line width to 0.75pt        

\begin{tikzpicture}[x=0.75pt,y=0.75pt,yscale=-1,xscale=1,line cap=round,line join=bevel]

%Shape: Rectangle [id:dp8885195769653287] 
\draw   (125,25) -- (225,25) -- (225,85) -- (125,85) -- cycle ;
%Shape: Ellipse [id:dp00032761180605045404] 
\draw   (30,57.99) .. controls (30,45.57) and (45.67,35.5) .. (65,35.5) .. controls (84.33,35.5) and (100,45.57) .. (100,57.99) .. controls (100,70.42) and (84.33,80.49) .. (65,80.49) .. controls (45.67,80.49) and (30,70.42) .. (30,57.99) -- cycle ;
%Straight Lines [id:da21852385987516665] 
\draw    (100,57.99) -- (123,58) ;
\draw [shift={(125,58)}, rotate = 180.01] [color={rgb, 255:red, 0; green, 0; blue, 0 }  ][line width=0.75]    (7.65,-2.3) .. controls (4.86,-0.97) and (2.31,-0.21) .. (0,0) .. controls (2.31,0.21) and (4.86,0.98) .. (7.65,2.3)   ;
%Shape: Rectangle [id:dp9916628671924079] 
\draw   (285,21) -- (480,21) -- (480,91) -- (285,91) -- cycle ;
%Straight Lines [id:da941480951298139] 
\draw    (225,60) -- (283,60) ;
\draw [shift={(285,60)}, rotate = 180] [color={rgb, 255:red, 0; green, 0; blue, 0 }  ][line width=0.75]    (7.65,-2.3) .. controls (4.86,-0.97) and (2.31,-0.21) .. (0,0) .. controls (2.31,0.21) and (4.86,0.98) .. (7.65,2.3)   ;
%Shape: Rectangle [id:dp9583945015361794] 
\draw   (285,105) -- (480,105) -- (480,140) -- (285,140) -- cycle ;
%Straight Lines [id:da053167742400991536] 
\draw    (385,90) -- (385,103) ;
\draw [shift={(385,105)}, rotate = 270] [color={rgb, 255:red, 0; green, 0; blue, 0 }  ][line width=0.75]    (7.65,-2.3) .. controls (4.86,-0.97) and (2.31,-0.21) .. (0,0) .. controls (2.31,0.21) and (4.86,0.98) .. (7.65,2.3)   ;
%Shape: Diamond [id:dp562167404459784] 
\draw   (385,156) -- (445,181) -- (385,206) -- (325,181) -- cycle ;
%Straight Lines [id:da9757195301238204] 
\draw    (385,140) -- (385,154) ;
\draw [shift={(385,156)}, rotate = 270] [color={rgb, 255:red, 0; green, 0; blue, 0 }  ][line width=0.75]    (7.65,-2.3) .. controls (4.86,-0.97) and (2.31,-0.21) .. (0,0) .. controls (2.31,0.21) and (4.86,0.98) .. (7.65,2.3)   ;
%Straight Lines [id:da9931852118292266] 
\draw    (384,205) -- (384,218) ;
\draw [shift={(384,220)}, rotate = 270] [color={rgb, 255:red, 0; green, 0; blue, 0 }  ][line width=0.75]    (7.65,-2.3) .. controls (4.86,-0.97) and (2.31,-0.21) .. (0,0) .. controls (2.31,0.21) and (4.86,0.98) .. (7.65,2.3)   ;
%Shape: Rectangle [id:dp3490288693127297] 
\draw   (341,220) -- (426,220) -- (426,245) -- (341,245) -- cycle ;
%Straight Lines [id:da6923000023240169] 
\draw    (325,181) -- (292,181) ;
\draw [shift={(290,181)}, rotate = 360] [color={rgb, 255:red, 0; green, 0; blue, 0 }  ][line width=0.75]    (7.65,-2.3) .. controls (4.86,-0.97) and (2.31,-0.21) .. (0,0) .. controls (2.31,0.21) and (4.86,0.98) .. (7.65,2.3)   ;
%Shape: Rectangle [id:dp25487748659624465] 
\draw   (180,160) -- (290,160) -- (290,205) -- (180,205) -- cycle ;
%Straight Lines [id:da05149667084436871] 
\draw    (240,160) -- (240,62) ;
\draw [shift={(240,60)}, rotate = 90] [color={rgb, 255:red, 0; green, 0; blue, 0 }  ][line width=0.75]    (7.65,-2.3) .. controls (4.86,-0.97) and (2.31,-0.21) .. (0,0) .. controls (2.31,0.21) and (4.86,0.98) .. (7.65,2.3)   ;

% Text Node
\draw (37,50.4) node [anchor=north west][inner sep=0.75pt]    {$\mathcal{X} ,\omega ({x})$};
% Text Node
\draw (176,55) node   [align=left] {\begin{minipage}[lt]{66.77pt}\setlength\topsep{0pt}
\begin{center}
Select a finite \\grid $\displaystyle \mathcal{X}_g \subset \mathcal{X}$;
\end{center}

\end{minipage}};
% Text Node
\draw (383,57) node   [align=left] [scale=1.2]{\begin{minipage}[lt]{150pt}\setlength\topsep{0pt}
\begin{center}
Solve \eqref{eq:discrete_constr} with $x \in \displaystyle \mathcal{X}_g$;\\ set $\displaystyle V({x}) \gets\textstyle\sum_{k=1}^N \alpha _{k} V_{k}({x})$
\end{center}

\end{minipage}};
% Text Node
\draw (380,123) node   [align=left] [scale=1.2]{ $\displaystyle {x}^{*} \gets\arg\underset{x\in \mathcal{X}}{\max}\, \Omega (\alpha,{x})$};
% Text Node
\draw (348,172.4) node [anchor=north west][inner sep=0.75pt]    {$\Omega \left(\alpha,{x}^{*}\right) \leq 0$};
% Text Node
\draw (392,203) node [anchor=north west][inner sep=0.75pt]   [align=left] {True};
% Text Node
\draw (383.5,233.5) node   [align=left] {Return $\displaystyle V({x})$};
% Text Node
\draw (294,161) node [anchor=north west][inner sep=0.75pt]   [align=left] {False};
% Text Node
\draw (185,175) node [anchor=north west][inner sep=0.75pt]    {$\mathcal{X}_g \gets\mathcal{X}_g \cup \left\{{x}^{*}\right\} ;$};
% Text Node
% \draw (151,67.4) node [anchor=north west][inner sep=0.75pt]    {$h\gets1;$};
% Text Node
% \draw (203,187.4) node [anchor=north west][inner sep=0.75pt]    {$h\gets h+1$;};

\end{tikzpicture}

%% file: pics/tikz/droneHier.tex
\tikzset{every picture/.style={line width=0.75pt}} %set default line width to 0.75pt        

\begin{tikzpicture}[x=0.75pt,y=0.75pt,yscale=-1,xscale=1, line join =bevel]

%Rounded Rect [id:dp7417278353201469] 
\draw  [color={rgb, 255:red, 42; green, 11; blue, 125 }  ,draw opacity=0.51 ][fill={rgb, 255:red, 255; green, 255; blue, 255 }  ,fill opacity=1 ] (178,83.15) .. controls (178,80.3) and (180.3,78) .. (183.15,78) -- (398.85,78) .. controls (401.7,78) and (404,80.3) .. (404,83.15) -- (404,142.85) .. controls (404,145.7) and (401.7,148) .. (398.85,148) -- (183.15,148) .. controls (180.3,148) and (178,145.7) .. (178,142.85) -- cycle ;
%Shape: Rectangle [id:dp35640252823670515] 
\draw  [draw opacity=0][fill={rgb, 255:red, 205; green, 108; blue, 108 }  ,fill opacity=0.34 ] (282,82) -- (400.5,82) -- (400.5,137.5) -- (282,137.5) -- cycle ;

%Shape: Rectangle [id:dp925219861959446] 
\draw  [fill={rgb, 255:red, 255; green, 255; blue, 255 }  ,fill opacity=1 ] (346,92) -- (398,92) -- (398,126) -- (346,126) -- cycle ;
%Image [id:dp6379811067949033] 
\draw (373.05,110.56) node  {\includegraphics[width=33.07pt,height=24.84pt]{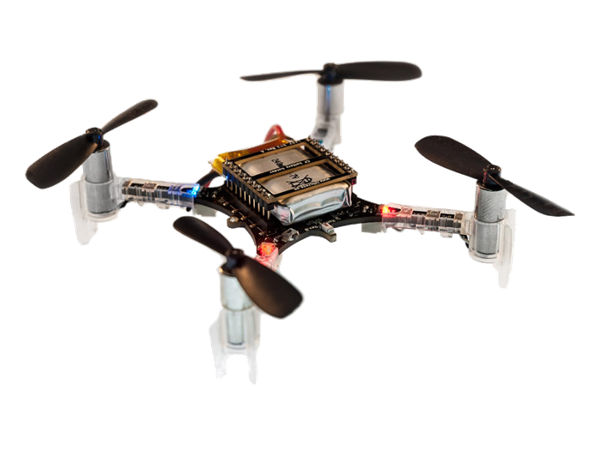}};
%Shape: Rectangle [id:dp5775024721767608] 
\draw  [fill={rgb, 255:red, 255; green, 255; blue, 255 }  ,fill opacity=1 ] (284,92) -- (332,92) -- (332,126) -- (284,126) -- cycle ;
%Straight Lines [id:da6089263780014667] 
\draw    (332,108) -- (344,108) ;
\draw [shift={(346,108)}, rotate = 180] [color={rgb, 255:red, 0; green, 0; blue, 0 }  ][line width=0.75]    (6.56,-1.97) .. controls (4.17,-0.84) and (1.99,-0.18) .. (0,0) .. controls (1.99,0.18) and (4.17,0.84) .. (6.56,1.97)   ;
%Straight Lines [id:da4597216930239527] 
\draw    (398,108) -- (408,108) -- (408,66) -- (320,66) ;
\draw [shift={(318,66)}, rotate = 360] [color={rgb, 255:red, 0; green, 0; blue, 0 }  ][line width=0.75]    (6.56,-1.97) .. controls (4.17,-0.84) and (1.99,-0.18) .. (0,0) .. controls (1.99,0.18) and (4.17,0.84) .. (6.56,1.97)   ;
%Straight Lines [id:da6425559551814] 
\draw    (370,126) -- (370,134) -- (306,134) -- (306,128) ;
\draw [shift={(306,126)}, rotate = 90] [color={rgb, 255:red, 0; green, 0; blue, 0 }  ][line width=0.75]    (4.37,-1.32) .. controls (2.78,-0.56) and (1.32,-0.12) .. (0,0) .. controls (1.32,0.12) and (2.78,0.56) .. (4.37,1.32)   ;
%Shape: Rectangle [id:dp7067184349017035] 
\draw  [fill={rgb, 255:red, 255; green, 255; blue, 255 }  ,fill opacity=1 ] (182,92) -- (246,92) -- (246,126) -- (182,126) -- cycle ;
%Straight Lines [id:da11385003138778371] 
\draw    (246,108) -- (282,108) ;
\draw [shift={(284,108)}, rotate = 180] [color={rgb, 255:red, 0; green, 0; blue, 0 }  ][line width=0.75]    (6.56,-1.97) .. controls (4.17,-0.84) and (1.99,-0.18) .. (0,0) .. controls (1.99,0.18) and (4.17,0.84) .. (6.56,1.97)   ;

%Straight Lines [id:da8541523762308287] 
\draw    (306,134) -- (212,134) -- (212,128) ;
\draw [shift={(212,126)}, rotate = 90] [color={rgb, 255:red, 0; green, 0; blue, 0 }  ][line width=0.75]    (4.37,-1.32) .. controls (2.78,-0.56) and (1.32,-0.12) .. (0,0) .. controls (1.32,0.12) and (2.78,0.56) .. (4.37,1.32)   ;
%Straight Lines [id:da02021398546874109] 
\draw    (280,62) -- (212,62) -- (212,90) ;
\draw [shift={(212,92)}, rotate = 270] [color={rgb, 255:red, 0; green, 0; blue, 0 }  ][line width=0.75]    (4.37,-1.32) .. controls (2.78,-0.56) and (1.32,-0.12) .. (0,0) .. controls (1.32,0.12) and (2.78,0.56) .. (4.37,1.32)   ;
%Straight Lines [id:da667758125080961] 
\draw    (344,58) -- (320,58) ;
\draw [shift={(318,58)}, rotate = 360] [color={rgb, 255:red, 0; green, 0; blue, 0 }  ][line width=0.75]    (6.56,-1.97) .. controls (4.17,-0.84) and (1.99,-0.18) .. (0,0) .. controls (1.99,0.18) and (4.17,0.84) .. (6.56,1.97)   ;

% Text Node
\draw (371,87.25) node  [scale=0.677] [align=left] {Crazyflie 2.1};
% Text Node
\draw (307.85,109) node  [scale=0.7]  [align=left] {\begin{minipage}[lt]{50pt}\setlength\topsep{0pt}
\begin{center}
Built-in\\inner loop
\end{center}

\end{minipage}};
% Text Node
\draw (214,109) node [scale=0.677] [align=left] {\begin{minipage}[lt]{45pt}\setlength\topsep{0pt}
\begin{center}
feedback\\linearization \\ \eqref{eq:linearization}
\end{center}

\end{minipage}};
% Text Node
\draw (247,95.5) node [anchor=north west][inner sep=0.75pt]  [font=\tiny]  {$F\leq \bar{F}$};
% Text Node
\draw (246,108.5) node [anchor=north west][inner sep=0.75pt]  [font=\tiny]  {$|\theta|  \leq \bar{\epsilon}$};

% Text Node
\draw (246,118.5) node [anchor=north west][inner sep=0.75pt]  [font=\tiny]  {$|\phi|  \leq \bar{\epsilon}$};

% Text Node
\draw (237,57) node  [font=\tiny]  {$u\in \mathcal{U}$ in \eqref{eq:Vctilde}};
%Shape: Rectangle [id:dp7082523889631098] 
\draw  [fill={rgb, 255:red, 191; green, 254; blue, 255 }  ,fill opacity=1 ] (267,48) -- (318,48) -- (318,73) -- (267,73) -- cycle ;
% Text Node
\draw (293.5,60) node  [scale=0.725] [align=left] {\begin{minipage}[lt]{50pt}\setlength\topsep{0pt}
\begin{center}
outer loop \\ $\bar u(x)$ in \eqref{control}
\end{center}

\end{minipage}};
% Text Node
\draw (181,138) node [anchor=north west][inner sep=0.75pt]  [font=\tiny] [align=left] {Linearized system $\ddot x_i = u_i $};
% Text Node
\draw (344,53) node [anchor=north west][inner sep=0.75pt]  [font=\tiny] [align=left] {Reference};

\end{tikzpicture}

%% file: pics/tikz/SCLF_3sys.tex
\tikzset{every picture/.style={line width=0.75pt}} %set default line width to 0.75pt        

\begin{tikzpicture}[x=0.75pt,y=0.75pt,yscale=-1,xscale=1, line join = bevel]
%uncomment if require: \path (0,353); %set diagram left start at 0, and has height of 353
%Image [id:dp13100223752200923] 
\draw (0,0) node  {\includegraphics[scale=0.4875]{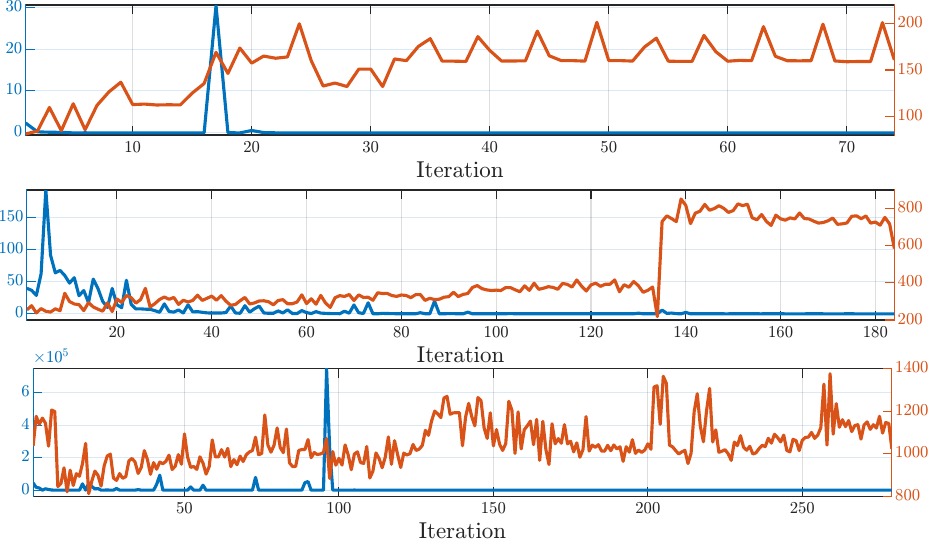}};

\draw (-160,-43) node [anchor=north west][inner sep=0.75pt, rotate=90, color={rgb, 255:red, 0; green, 114; blue, 189 }]  [font=\tiny]  {$\max\Omega(\cdot)$};

\draw (-160,18.5) node [anchor=north west][inner sep=0.75pt, rotate=90, color={rgb, 255:red, 0; green, 114; blue, 189 }]  [font=\tiny]  {$\max\Omega(\cdot)$};

\draw (-160,70) node [anchor=north west][inner sep=0.75pt, rotate=90, color={rgb, 255:red, 0; green, 114; blue, 189 }]  [font=\tiny]  {$\max\Omega(\cdot)$};

\draw (148,-49) node [anchor=north west][inner sep=0.75pt, rotate=90, color={rgb, 255:red, 216; green, 83; blue, 25 }]  [font=\tiny]  {BT (s)};

\draw (148,10) node [anchor=north west][inner sep=0.75pt, rotate=90, color={rgb, 255:red, 216; green, 83; blue, 25 }]  [font=\tiny]  {BT (s)};

\draw (148,62) node [anchor=north west][inner sep=0.75pt, rotate=90, color={rgb, 255:red, 216; green, 83; blue, 25 }]  [font=\tiny]  {BT (s)};

\end{tikzpicture}

%% file: pics/tikz/bioreact_sim.tex
\tikzset{every picture/.style={line width=0.75pt}} %set default line width to 0.75pt        

\begin{tikzpicture}[x=0.75pt,y=0.75pt,yscale=-1,xscale=1]
%uncomment if require: \path (0,353); %set diagram left start at 0, and has height of 353
%Image [id:dp13100223752200923] 
\draw (0,0) node  {\includegraphics[scale=0.6]{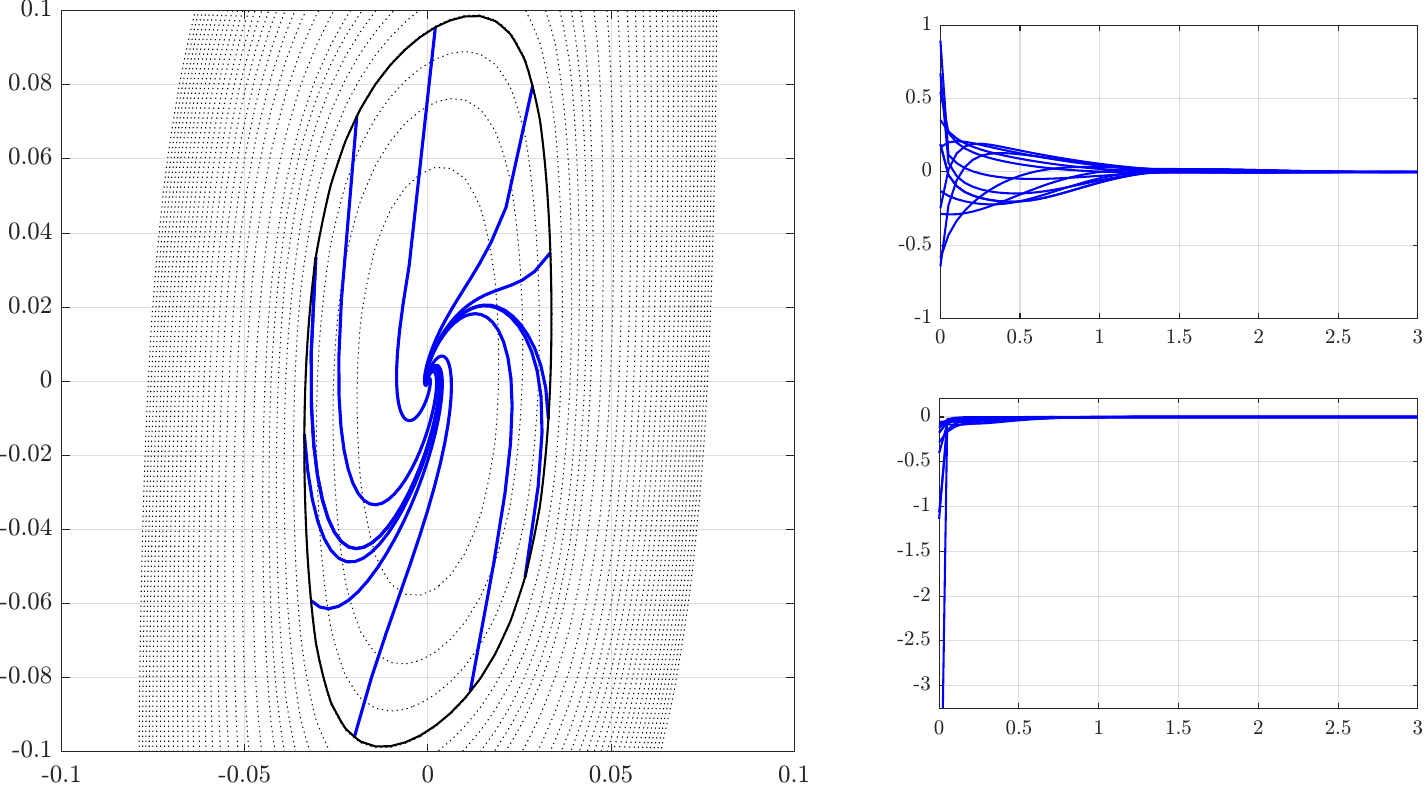}};
\draw (-125,158) node [anchor=north west][inner sep=0.75pt]  [font=\large]  {$x_1$};
\draw (-290,8) node [anchor=north west][inner sep=0.75pt, rotate=90]  [font=\large]  {$x_2$};

% \draw(26,100) node [anchor=north west][inner sep=0.75pt, rotate=90]  [font=\small]  {CT (ms)};

\draw (50,95) node [anchor=north west][inner sep=0.75pt, rotate=90]  [font=\normalsize]  {$\Theta({x})$ in \eqref{eq:thetaSum}};

\draw (50,-70) node [anchor=north west][inner sep=0.75pt, rotate=90]  [font=\normalsize]  {$\bar u(x)$};

\draw (155,135) node [anchor=north west][inner sep=0.75pt]  [font=\normalsize]  {time $t$};

\end{tikzpicture}

%% file: pics/tikz/vtol_sim_1col_clean.tex
\tikzset{every picture/.style={line width=0.75pt}} %set default line width to 0.75pt        

\begin{tikzpicture}[x=0.75pt,y=0.75pt,yscale=-1,xscale=1, line join = bevel]
%uncomment if require: \path (0,353); %set diagram left start at 0, and has height of 353
%Image [id:dp13100223752200923] 
\draw (0,0) node  {\includegraphics[scale=0.4875]{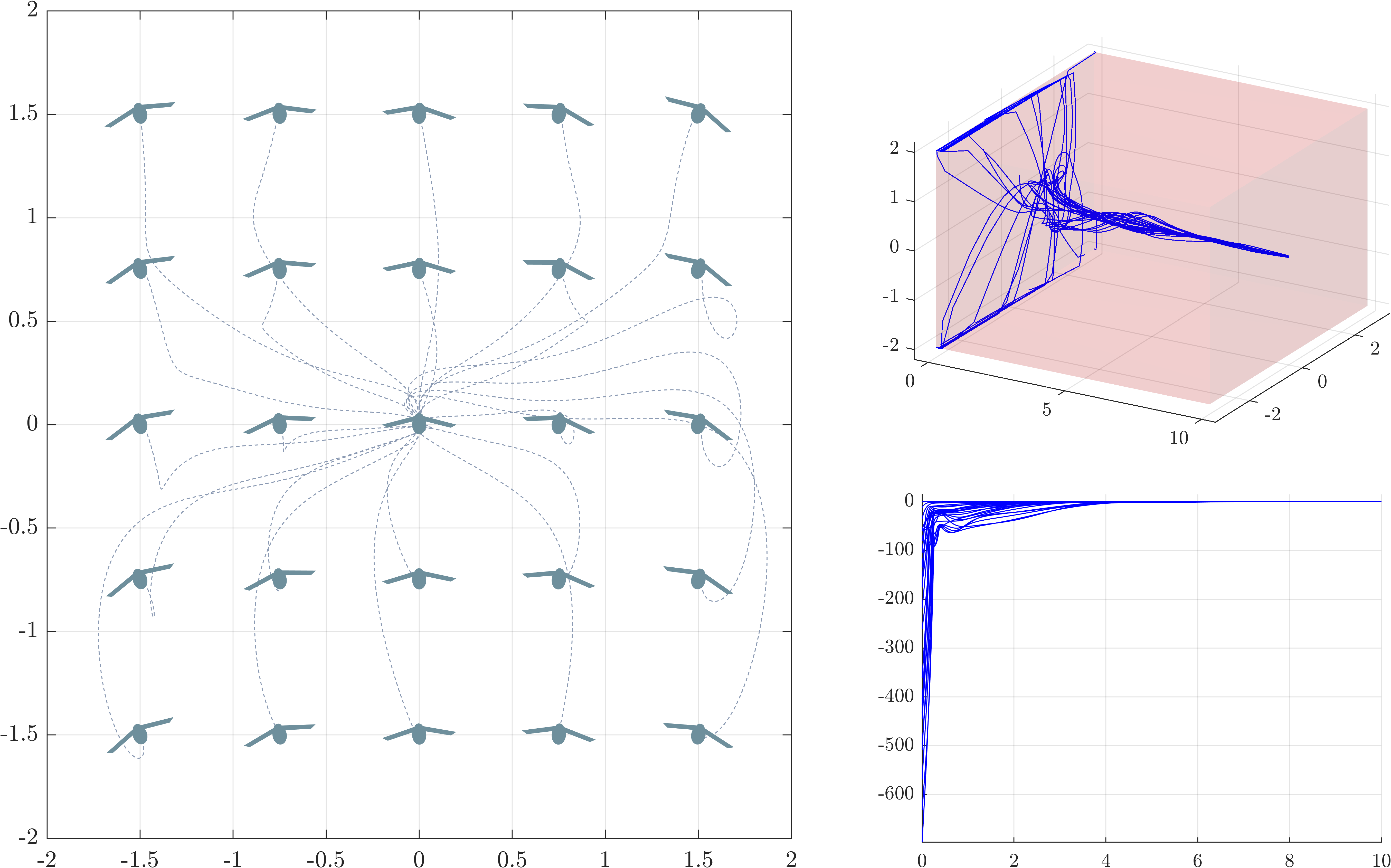}};

\draw (-100,152) node [anchor=north west][inner sep=0.75pt]  [font=\normalsize]  {$x_1$};
\draw (-243,10) node [anchor=north west][inner sep=0.75pt, rotate=90]  [font=\normalsize]  {$x_2$};

% \draw(8,132) node [anchor=north west][inner sep=0.75pt, rotate=90]  [font=\tiny]  {CT (ms)};

\draw (39,109) node [anchor=north west][inner sep=0.75pt, rotate=90]  [font=\small]  {$\Theta(x)$ in \eqref{eq:thetaSum}};

\draw (45,-60) node [anchor=north west][inner sep=0.75pt, rotate=90]  [font=\small]  {$u_2$};
\draw (212,-16) node [anchor=north west][inner sep=0.75pt]  [font=\small]  {$u_1$};
\draw (72.5,-3) node [anchor=north west][inner sep=0.75pt]  [font=\small]  {time (s)};

\draw (130,148) node [anchor=north west][inner sep=0.75pt]  [font=\small]  {time (s)};
% \draw (31,121) node [anchor=north west][inner sep=0.75pt]  [font=\tiny]  {time (s)};
%Curve Lines [id:da6926053506441407] 
% \draw  [shift={(80,-150-80-15-35)}]    (80,130) .. controls (66.53,116.67) and (42.06,122.55) .. (40.09,148.39) ;
% \draw [shift={(120,150-150-80-15-35)}, rotate = 272.22] [color={rgb, 255:red, 0; green, 0; blue, 0 }  ][line width=0.75]    (4.37,-1.32) .. controls (2.78,-0.56) and (1.32,-0.12) .. (0,0) .. controls (1.32,0.12) and (2.78,0.56) .. (4.37,1.32)   ;
% \draw (140,-151) node [anchor=north west][inner sep=0.75pt,fill=white,draw ={rgb, 255:red, 0; green, 0; blue, 0 },line width = 0.1]  [font=\tiny]  {input constraint set};

\end{tikzpicture}